\documentclass[showpacs,preprintnumbers,amsmath,amssymb,aip,pof,graphicx]{revtex4}

\usepackage{amssymb,amsfonts,amsmath}
\usepackage{euscript}
\usepackage{graphics}
\usepackage{setspace}
\usepackage{graphicx}
\usepackage{hyperref}

\begin{document}


\title{Auto-chemotactic micro-swimmer suspensions: modeling, analysis and simulations} 



\author{Enkeleida Lushi$^{1,2}$, Raymond E. Goldstein$^{3}$, Michael J. Shelley$^1$}
\affiliation{$^1$Courant Institute of Mathematical Sciences, New York University, New York, NY 10012, USA\\
$^2$School of Engineering, Brown University, Providence, RI, 02912, USA\\
$^3$Department of Applied Mathematics and Theoretical Physics, University of Cambridge, Cambridge CB3 0WA, United Kingdom}





\begin{abstract}
  Microorganisms can preferentially orient and move along gradients of
  a chemo-attractant (i.e., chemotax) while colonies of many
  microorganisms can collectively undergo complex dynamics in response
  to chemo-attractants that they themselves produce. For colonies or
  groups of micro-swimmers we investigate how an ``auto-chemotactic''
  response that should lead to swimmer aggregation is affected by the
  non-trivial fluid flows that are generated by collective
  swimming. For this, we consider chemotaxis models based upon a
  hydrodynamic theory of motile suspensions that are fully coupled to
  chemo-attractant production, transport, and diffusion. Linear
  analysis of isotropically ordered suspensions reveals both an
  aggregative instability due to chemotaxis that occurs independently
  of swimmer type, and a hydrodynamic instability when the swimmers
  are ``pushers''. Nonlinear simulations show nonetheless that
  hydrodynamic interactions can significantly modify the
  chemotactically-driven aggregation dynamics in suspensions of
  ``pushers'' or ``pullers''.  Different states of the dynamics
  resulting from these coupled interactions in the colony are
  discussed.
\end{abstract}

\keywords{chemotaxis, locomotion, cell motility, hydrodynamics, kinetic theory, microorganisms, suspensions }

\pacs{87.17.Jj, 05.20.Dd, 47.63.Gd, 87.18.Hf}

\maketitle

\section{Introduction}
\label{Introduction}

Recent advances in experiment and in theoretical and computational
modeling have established that suspensions of motile microorganisms
can organize into complex patterns and collectively generate
significant fluid flows \cite{ CisnEtAl07, DombEtAl04, SaintShelley08,
  SaintShelley08b, SubKoch09}. These large-scale patterns can occur in
the bulk in the absence of directional cues for swimming and are
mediated by steric and/or hydrodynamic interactions between the
micro-swimmers \cite{Dunkel13, Hensink12, SokolovAranson12}. 
It is also well-known that motile microorganisms can
exhibit directed chemotactic motions in response to chemical cues in
their environment. When those cues are attractive and produced by the
motile organisms themselves, then collective aggregation can occur. We
refer to such a situation as ``auto-chemotactic'' in that the colony is
responding to its own self-generated signals. However, many of the
classical experiments on auto-chemotactic aggregation, which can show
intricate patterns such as stripes, fingers, and arrays of spots, were
performed in environments where hydrodynamic coupling between the
motile cells is not expected to be strong (e.g. in the thin fluid
layer atop an agar plate \cite{BudreneBerg, BudreneBerg2}).
Auto-chemotactic systems are considerably more complicated when the
constituent organisms are moving in an open fluid and can generate
collective flows since these flows will also advect the
chemoattractant \cite{Leptos09}. Hence it is natural to ask how
collectively generated flows in suspensions of microorganisms
\cite{Rushkin10} affect chemotactic aggregation and patterning, and
hence possibly affect modes of colonial communication such as through
quorum sensing \cite{Bassler02,ParkEtAl03}. Here we investigate these
issues through a theoretical model that combines the collective flows
generated by a motile suspension, with the production, advection, and
diffusion of a swimmer-generated chemo-attractant, and the response of
the swimmers to the chemo-attractant field.

Theoretically, chemotactic aggregation and pattern formation has been
studied extensively using the Keller-Segel (KS) model
\cite{KellerSegel70,KellerSegel71} and its many variants. The KS model
couples evolution of a cell concentration field to an intrinsically
generated, diffusing chemo-attractant field. In its barest form, where
the cell velocity scales linearly with chemoattractant gradient, the
KS model can lead to infinite concentrations in finite time
\cite{Childress84,DolPert04,JagerLuck92}.  Such behaviors can be
avoided through the inclusion of {\it ad hoc} saturation
terms (see Tindal {\em et al.} \cite{TindalEtAl08} for a review).

More recently, kinetic theories have been developed for the dynamics
of bacterial populations where the individual organisms are executing
modulated run-and-tumble motions in response to a chemoattractant
gradient
\cite{Alt80,BearonPedley00,Schnitzer93,TindalEtAl08,SaragostiEtAl11}.
In these models, tumbling frequency decreases (and hence run length
increases) if the organism if moving up the attractant gradient as is
observed experimentally \cite{BergBrown72, Berg93}. In these models
swimmer speeds are bounded and so no singular behavior is expected (at
least in finite time).

Here, as in our recent preliminary study \cite{LGS12}, we consider
here a kinetic model for modulated run-and-tumble chemotactic dynamics
that includes the effect of fluid flows produced by collective
swimming. These fluid flows will both advect any chemoattractant, and
perturb the motions of the constituent swimmers. A simpler version of
our extended model was considered by Bearon \& Pedley
\cite{BearonPedley00} who studied chemotaxis in a given background
shear flow. Without the run-and-tumble dynamics, our model reduces to
the kinetic model for active suspensions developed by Saintillan \&
Shelley \cite{SaintShelley08, SaintShelley08b}, which captures the
large-scale swimmer-induced flows observed in experiments
\cite{CisnEtAl07,CisnEtAl11, DombEtAl04}, and which illuminates the
effect of the propulsive mechanism (pusher vs. puller) and swimmer
shape. Merging the the run-and-tumble and the active suspension models
is seamless and natural as both are kinetic theories with particle
position and orientation as their conformation variables
\cite{LushiDissert,LGS12}.

We first study the linear stability of steady-state, isotropic
suspensions (that is, uniform in density and orientation) for a
simplified version of the extended model. The linearized system yields
two separated branches of instability, one associated with
chemotactically driven aggregation, and a ``hydrodynamic'' instability
that drives swimmer alignment through the development of large-scale
fluid flows. The latter is a feature of active ``pusher'' suspensions, has been
extensively analyzed 
\cite{SaintShelley08b, SubKoch09, HohenShelley10}, and is not
present when the swimming particles are ``pullers''. Our analysis also
identifies regions of parameter space where the hydrodynamic and
aggregative instabilities are separately and jointly important. The
ultimate state of the fully coupled model is studied through nonlinear
simulation, which shows that swimmer generated fluid flows can have a
profound effect on aggregation dynamics. In particular, Pusher
suspensions can create complex flows that move and fragment
concentrated regions of swimmers, and so bound growth in swimmer
density, while Puller suspensions evolve into circular aggregates that
mutually repel each other through their intrinsically generated flows,
thereby limiting aggregate coarsening.

As an interesting alternative to modulated run-and-tumble chemotaxis,
we also analyze and simulate a different model wherein the constituent
swimmers chemotax by directly detecting spatial chemo-attractant
gradients and responding to it by biasing their
direction. This kind of model is more appropriate for
larger swimmers such as eukaryotic spermatozoa. Despite its
differences, we find that this ``turning-particle'' chemotaxis model
exhibits many of the same dynamical features as the run-and-tumble
model.

\section{Mathematical Model}
\label{Models}

\subsection{The Run-and-Tumble Model}

We first consider how a \textit{Run-and-Tumble} (RT) chemotactic
response can be incorporated into a kinetic theory of motile
suspensions. Bacteria such as \textit{Escherichia coli} are known to
perform a biased random walk which enables them to move, on average, up
chemo-attractant gradients \cite{Berg93}. Such a random walk consists
of a series of runs and tumbles whose frequency decreases when a
bacterium is moving in a direction of increasing chemo-attractant
concentration. The RT chemotaxis model we use here is based on Alt's
classical work \cite{Alt80}, and extends the models of Schnitzer
\cite{Schnitzer93}, Bearon and Pedley \cite{BearonPedley00} and Chen
\textit{et.al.} \cite{ChenEtAl03} on a continuum formulation of the
biased random walk in three dimensions.

Consider $N$ self-propelled ellipsoidally-shaped swimmers each moving
with intrinsic speed $U_0$ in a box-shaped fluid domain of dimension
$L$ and volume $V=L^3$. The swimmer center-of-mass is denoted by
$\mathbf{x}$ with the swimming direction $\mathbf{p}$ (with
$\mathbf{p}\cdot\mathbf{p}=1$) along its main axis. We represent the
configuration of micro-swimmers by a distribution function
$\Psi(\mathbf{x},\mathbf{p},t)$. The positional and orientational
dynamics of a suspension of swimmers that individually execute a
run-and-tumble dynamics is described by a 
Fokker-Planck equation for conservation of particle number:
\begin{align}
 \frac{\partial \Psi}{\partial t} &= - \nabla_x \cdot [ \Psi \dot{\mathbf{x}} ] - \nabla_p \cdot [ \Psi   \dot{ \mathbf{p}} ] \nonumber \\
 &- [\Psi \lambda (\mathcal{D}_t C)  - \frac{1}{4\pi}\int \Psi  (\mathbf{p'}) \lambda (\mathcal{D}_t C)  d \mathbf{p'}] \label{runandtumble3D} \\
  \dot{\mathbf{x}} &= U_0 \mathbf{p} + \mathbf{u}  - D \nabla_x (\ln \Psi)  \label{xdot} \\
  \dot{ \mathbf{p}} &=  (\mathbf{I}-\mathbf{pp})(\gamma \mathbf{E} + \mathbf{W}) \mathbf{p} - d_r \nabla_p (\ln \Psi). \label{pdot}
\end{align}
Equations~(\ref{xdot}) and (\ref{pdot}) give the conformational fluxes
associated with swimmer position and orientation. Equation
(\ref{xdot}) states that a particle propels itself along its axis
$\mathbf{p}$ with speed $U_0$ while being carried along in the
background flow $\mathbf{u}$. The last term in the flux allows for an
isotropic translational diffusion with diffusion constant
$D$. Equation (\ref{pdot}) is Jeffery's equation \cite{Jeffrey22} for
the rotation of an ellipsoidal particle by the local background flow,
with $\mathbf{E} = (\nabla \mathbf{u} + \nabla^T \mathbf{u} )/2$,
$\mathbf{W} = (\nabla \mathbf{u} -\nabla^T \mathbf{u} )/2$ the
rate-of-strain and vorticity tensors, respectively, and $\gamma$ a
shape parameter $-1 \leq \gamma \leq 1$ (for an ellipsoidal particle
with aspect ratio A, $\gamma = (A^2-1)/(A^2+1)$; for a sphere
$\gamma=0$ and for a slender rod $\gamma \approx 1$). The last term,
with $\nabla_{\mathbf{p}}$ being the gradient operator on the sphere
$|\mathbf{p}|=1$, models rotational diffusion of the swimmer with
diffusion constant $d_r$, as in \cite{SaintShelley08b}.

Run-and-tumble chemotaxis, based on straight runs and modulated
tumbles, is modeled by the last term (in brackets) of
Eq.~(\ref{runandtumble3D}). The first part is a loss term of swimmers
tumbling from orientation $\mathbf{p}$ to other orientations, and the
second term, a nonlocal flux, is a balancing source to account for
swimmers tumbling from other orientations $\mathbf{p'}$ to
$\mathbf{p}$. Here, $\lambda(\mathcal{D}_t C)$ is the chemical
gradient-dependent tumbling frequency with $C(\mathbf{x}, t )$ being
the chemo-attractant concentration.  The tumbling frequency is related
to the probability of a bacterium having a tumbling event within a
fixed time interval.

From experimental observation \cite{MacnabKoshland72}, when the time
rate-of-change of the chemo-attractant concentration is positive along
the path of a swimmer, the swimmer's tumbling rate reduces. If the
chemo-attractant concentration is constant or decreasing, the tumbling
rate is constant. Based on experimental data \cite{MacnabKoshland72}
and theoretical studies, such as \cite{ChenEtAl03}, this biphasic
response for $\lambda(\mathcal{D}_t C)$ has been modeled as
\begin{eqnarray}\label{stoppingrate}
 \lambda(\mathcal{D}_t C) = \left\{
  \begin{array}{l l}
     \lambda_0 \exp \left( -\chi \mathcal{D}_t C  \right) & \quad \text{if } \mathcal{D}_t C  >0\\
    \lambda_0 & \quad \text{otherwise},
  \end{array} \right.
\end{eqnarray}
where
\begin{align} \label{DcDt} \mathcal{D}_t C 
= \frac{\partial C}{\partial t} 
+ \left( \mathbf{u} + U_0 \mathbf{p} \right) \cdot \nabla{C}
\end{align}
is the rate-of-change of the chemo-attractant concentration along the
bacterium path. The parameter $\lambda_0$ is the basal stopping rate,
or tumbling frequency, in the absence of chemotaxis, and $\chi$ the
chemotactic strength. In the literature the frequency response
$\lambda$ has been approximated in various forms, the exponential as
above \cite{ChenEtAl03} or by a linearized form \cite{BearonPedley00},
and often does not include the temporal gradient
\cite{TindalEtAl08}. These studies do not include chemo-attractant
dynamics or hydrodynamics.

Here we model the tumbling frequency $\lambda(\mathcal{D}_t C)$ using a
simple piece-wise linearized form of Eq. (\ref{stoppingrate})
\begin{eqnarray}\label{stoppingrate_lin}
 \lambda(\mathcal{D}_t C) = \left\{
  \begin{array}{l l}
     \lambda_0 \left( 1 -\chi \mathcal{D}_t C \right)& \quad \text{if } 0< \mathcal{D}_t C <1/\chi \\
     0 & \quad \text{if } 1/\chi< \mathcal{D}_t C\\
    \lambda_0 & \quad \text{otherwise}
  \end{array} \right.
\end{eqnarray}

It is possible to include an anisotropic tumbling in the integral term
in Eq. (\ref{runandtumble3D}) via a ``turning kernel'' that is
dependent on $|\mathbf{p} - \mathbf{p'}|$, where $\mathbf{p}$ and
$\mathbf{p'}$ are pre- and post-tumble directions (E. Lushi, in
preparation). In this paper we focus on isotropic tumbles only.

The fluid velocity $\mathbf{u}(\mathbf{x},t)$ satisfies the Stokes equations 
with an extra stress due to the particles' motion in it,
\begin{align}\label{Stokes-dim}
 -\mu \nabla_x^2 \mathbf{u} +\nabla_x q &= \nabla_x \cdot \Sigma^a \nonumber \\
 \nabla_x \cdot \mathbf{u} &= 0.
\end{align}
Here $\mu$ is the immersing fluid's viscosity, $q$ the fluid pressure,
$\Sigma^a$ the active particle stress derived using Kirkwood theory
\cite{DoiEdwards86},
\begin{align}\label{stress-nondim}
 \Sigma^a (\mathbf{x},t) = \sigma_0 \int \Psi (\mathbf{x},\mathbf{p},t) (\mathbf{pp}^T-\mathbf{I}/3)d\mathbf{p}.
\end{align}
The active stress $\Sigma^a$ is a configuration average over all orientations 
$\mathbf{p}$ of the stresslets $\sigma_0 (\mathbf{pp}^T-\mathbf{I}/3)$ exerted by 
the particles on the fluid. The stresslet strength $\sigma_0$ arises from the first moment 
of the force distribution on the particle surface, where particle interactions are neglected 
and only the lowest order contribution from single-particle swimming is retained 
\cite{SaintShelley08b}. It can be shown from the swimming micro-mechanics that
\begin{align}\label{alph}
 \sigma_0 = U_0 \mu l^2 \alpha,
\end{align}
where $l$ is the characteristic length of a particle and $\alpha$ is a $O(1)$ 
dimensionless constant that depends on the mechanism of swimming and swimmer 
geometry. For \textit{pushers}, swimmers that propel themselves by exerting a force 
near the tail, e.g. the bacteria \textit{B. subtilis} or \textit{E. coli}, we have $\sigma_0<0$ 
and thus $\alpha<0$.  For \textit{pullers}, swimmers that propel themselves from the 
head, e.g. the biflagellated alga \textit{Chlamydomonas reinhardtii}, we have $\sigma_0>0$ and $\alpha>0$.

We define a local particle/swimmer concentration $\Phi (\mathbf{x},t)$ by 
\begin{align}\label{PhiEqn}
 \Phi(\mathbf{x},t) &= \int \Psi (\mathbf{x},\mathbf{p},t) d \mathbf{p}, 
\end{align}

The chemo-attractant or nutrient is also dispersed in the fluid and has a dynamics 
of its own that includes fluid advection and molecular diffusion. Similar to the original 
KS model \cite{KellerSegel71} but with fluid advection included, the 
chemo-attractant dynamics obeys
\begin{align}\label{chemo}
 \frac{\partial C}{\partial t} + \mathbf{u} \cdot \nabla C = -\beta_1 C +\beta_2 \Phi + D_c \nabla^2 C.
\end{align}
where $-\beta_1 C$ models chemo-attractant degradation with constant rate $\beta_1$, and $\beta_2 \Phi$ 
describes local production ($\beta_2>0$) or consumption ($\beta_2<0$) of 
chemo-attractant by the micro-swimmers. The last term models spatial diffusion with 
diffusion coefficient $D_c$. We differentiate between two types of chemotaxis: 
in \textit{auto-chemotaxis}, for $\beta_2>0$, the swimmers produce the 
chemo-attractant, whereas in \textit{oxygen-taxis} the micro-swimmers respond 
and consume a chemo-attractant (e.g. a nutrient like oxygen) which is 
externally-supplied ($\beta_2 \leq 0$). The focus of this paper is on auto-chemotaxis, whereas the 
oxygen-taxis aspect is investigated in \cite{LushiDissert} and more 
recently in \cite{Ezhilan12}.

Taken together, the chemo-attractant equation (\ref{chemo}), the equation (\ref{runandtumble3D}) 
for the probability distribution function $\Psi$ (and hence $\Phi$) and the Stokes 
equations (\ref{Stokes-dim}) with active particle stress, constitute a closed system that 
describes the dynamics of a motile suspension influenced by run-and-tumble 
chemotaxis in an evolving chemical field. We will refer to this model as the 
\textit{Run-and-Tumble Chemotaxis} model.

\subsection{The Turning-Particle Model}

As an interesting alternative, we also consider a generalized
chemotaxis model for suspensions of non-tumbling micro-swimmers that
can directly measure a chemo-attractant gradient and respond to
it. This kind of model is appropriate for larger swimmers, such as
eukaryotic spermatozoa.

The configuration of micro-swimmers is again represented by a
distribution function $\Psi(\mathbf{x},\mathbf{p},t)$ of the center of
mass position $\mathbf{x}$ and orientation vector $\mathbf{p}$. The
dynamics is described by the conservation equation
\begin{align}
 \frac{\partial \Psi}{\partial t} &= - \nabla_x \cdot [ \Psi \dot{\mathbf{x}}  ] - \nabla_p \cdot [ \Psi \dot{\mathbf{p}}  ] \label{Psi_cons_eqn}\\
  \dot{\mathbf{x}} &= U_0 \mathbf{p} + \mathbf{u}- D \nabla_x (\ln \Psi) \label{xdot_tp} \\
  \dot{\mathbf{p}} &= (I- \mathbf{p}\mathbf{p}^{ T}) \left[ (\gamma \mathbf{E} + \mathbf{W}) \mathbf{p} + \xi \nabla_x C\right] - d_r \nabla_p(\ln{\Psi}) \label{pdot_tp}
\end{align}
While there are now no tumbling related terms in
Eq. \ref{Psi_cons_eqn}, in Eq. \ref{pdot_tp} there is now a term,
$\xi(\mathbf{I}- \mathbf{p}\mathbf{p}^T) \nabla C$ which induces a
``chemotactic'' swimmer rotation towards the local direction of
steepest ascent in chemo-attractant concentration. The constant $\xi$
helps set the time-scale of this rotation. This rotation is differentiated from
rotational diffusion, which acts on very rapid time-scales and is
associated with very small changes in direction.

The chemo-attractant equation (\ref{chemo}), together with the
equation (\ref{Psi_cons_eqn}) for the probability distribution
function $\Psi$, and the Stokes equations (\ref{Stokes-dim}) with
active particle stress, make a closed system of equations that
describes the dynamics of a chemotactic motile suspension with an
evolving chemical field. We will refer to this set of equations as the
\textit{Turning-Particle Chemotaxis} model.

\subsection{A Note on Non-Dimensionalization}

We make Eqs. (\ref{runandtumble3D}-\ref{pdot_tp}) non-dimensional by
rescalings based on the swimmer contribution to the fluid stress
tensor \cite{SaintShelley08b}. As a characteristic length we use
$\ell=l/\nu$, where $l$ is the swimmer size and $\nu \equiv Nl^3/V$ is
the effective mean volume fraction of swimmers. Velocity and time are
re-scaled by $U_0$ and $l/U_0$, respectively, and the distribution
function normalized so that
\begin{align*}
 \frac{1}{V} \int_V d\mathbf{x} \int d\mathbf{p} \Psi(\mathbf{x}, \mathbf{p}, t) = 1.
\end{align*}
Consequently, $\Psi_0\equiv
1/4\pi$ is the probability density for the uniform isotropic
state. Under these choices, the swimming speed is unity, the viscosity
is unity in the Stokes equation (\ref{Stokes-dim}), and $\sigma_0$ is replaced
by $\alpha$ [see Eq. (\ref{alph})]. Recall that $\sigma_0$, and hence
$\alpha$, is signed (Pushers: $alpha<0$; Pullers: $\alpha>0$). We will
also consider the case of ``neutral swimmers'', for which $\alpha=0$
but the swimming speed remains unity (i.e. no hydrodynamic interactions).
Note that the swimmer volume fraction $\nu$ now appears in the
rescaled system size $L/\ell$.

\section{Stability Analysis}
\label{Stability}

\subsection{Linear Stability of Run-and-Tumble Auto-Chemotaxis}
\label{3Dlinearanalysis}
\subsubsection{The Eigenvalue Problem}

We analyze the linear stability of auto-chemotactic uniform isotropic
suspensions, $\beta_1,\beta_2>0$ in Eq. (\ref{chemo}). For simplicity,
we consider only a quasi-static chemo-attractant field,
\begin{align}\label{quasistaticchem}
 - \beta_1 C + \beta_2 \Phi+ D_c \nabla_x^2 C =0.
\end{align}
and no translational or rotational diffusion $D=d_r=0$.

We consider plane-wave perturbations of the distribution and 
chemo-attractant concentration functions about the uniform isotropic 
state ($\Psi_0=1/4\pi$) and steady-state ($\overline{C} = \beta_2/\beta_1$),
respectively:
\begin{eqnarray}\label{Psi_perturb_rt}
\Psi(\mathbf{x}, \mathbf{p},t) &= 1/4\pi + \epsilon \tilde{\Psi}(\mathbf{p},\mathbf{k}) \exp (i \mathbf{k} \cdot \mathbf{x} + \sigma t) \nonumber \\
C(\mathbf{x}, t) &= \beta_1/ \beta_2 + \epsilon \tilde{C}(\mathbf{k})  \exp (i \mathbf{k} \cdot \mathbf{x} + \sigma t), \nonumber 
\end{eqnarray}
with $|\epsilon|\ll 1$, $\mathbf{k}$ the wavenumber and $\sigma$ the growth 
rate. The tumbling frequency is then simplified to
\begin{align}
\lambda(\mathcal{D}_t C) = \lambda_0\left(1- \chi \mathbf{p} \cdot \nabla C  \right).
\end{align}

Retaining only first-order terms in $\epsilon$ we find
\begin{align}\label{tildePsi}
\sigma \tilde{\Psi} = &-i\mathbf{p} \cdot \mathbf{k} \tilde{\Psi} +
\frac{3 \gamma}{4\pi} \mathbf{pp} : \tilde{\mathbf{E}} \nonumber
\\ 
&-\lambda_0 \tilde{\Psi} + \frac{\lambda_0 \chi}{4 \pi} \mathbf{p}
\cdot \mathbf{k} \tilde{C} + \frac{\lambda_0 }{4 \pi} \int
\tilde{\Psi'} d\mathbf{p'},
\end{align}
and $\tilde{C}$ is given in terms of $\tilde{\Psi}$ by
\begin{align}\label{tildeC}
\tilde{C} = \frac{\beta_2}{\beta_1 + k^2 D_c}\tilde{\Phi} = 
\frac{\beta_2}{\beta_1 + k^2 D_c}\int \tilde{\Psi'} d\mathbf{p'},
\end{align}
where $k= |\mathbf{k}|$. The fluid velocity perturbation can be
expressed in terms of the active stress perturbation by
\begin{align}\label{vel_lin}
\tilde{\mathbf{u}} = \frac{i}{k}(\mathbf{I} - \hat{\mathbf{k}}\hat{\mathbf{k}}^T) \cdot \tilde{\Sigma^a} \cdot \hat{\mathbf{k}},
\end{align}
where
\begin{align}\label{stress_lin}
\tilde{\Sigma^a} = \alpha \int \tilde{\Psi'} \mathbf{p'p'}^T d\mathbf{p'}.
\end{align}
The perturbed symmetric rate-of-strain tensor is
\begin{align}
\tilde{\mathbf{E}} &= \frac{i}{2}(\tilde{\mathbf{u}}\mathbf{k}^T +
\mathbf{k} \tilde{\mathbf{u}}^T ), \nonumber \\
&= -\alpha(\mathbf{I} -
\hat{\mathbf{k}}\hat{\mathbf{k}}^T) \cdot \int \tilde{\Psi'}
\mathbf{p'p'}^T d\mathbf{p'} \cdot \hat{\mathbf{k}}\hat{\mathbf{k}}^T.\label{tildeE}
\end{align}
Substituting Eqs. (\ref{tildeC}) \& (\ref{tildeE}) into
Eq. (\ref{tildePsi}), we arrive at the eigenvalue/eigenmode relation
\begin{align}\label{PsiLinDecomp}
&(\sigma + \lambda_0 + i \mathbf{p} \cdot \mathbf{k} ) \tilde{\Psi}
  \nonumber \\ &= -\frac{3 \alpha \gamma}{4 \pi}(\hat{\mathbf{k}}
  \cdot \mathbf{p})
  \mathbf{p}^T(\mathbf{I}-\hat{\mathbf{k}}\hat{\mathbf{k}}^T) \int
  \tilde{\Psi'} \mathbf{p'}(\mathbf{p'}\cdot \mathbf{\hat{k}})
  d\mathbf{p'} \nonumber \\ &+ \frac{\lambda_0}{4 \pi} \left(
  \frac{ik\chi \beta_2 }{\beta_1 + k^2 D_c}(\hat{\mathbf{k}} \cdot
  \mathbf{p}) + 1 \right) \int \tilde{\Psi'} d\mathbf{p'}.
\end{align} 
The first term on the right hand side of Eq.(\ref{PsiLinDecomp}) is
the hydrodynamic coupling term that resulting from inverting the Stokes 
Equations and its action is restricted to the second azimuthal mode on
$|\mathbf{p}|=1$ (see also Hohenegger \& Shelley \cite{HohenShelley10}). The
second term on the right hand side of Eq.(\ref{PsiLinDecomp}) 
is the auto-chemotactic term resulting from inverting the 
quasi-static chemo-attractant Eq. (\ref{quasistaticchem}), and its dynamics is
restricted to the zeroth azimuthal mode on $|\mathbf{p}|=1$ (see Lushi
{\it et. al.} \cite{LGS12}). To show this clearly, we define the
operators $\mathbf{F}$ and $\mathbf{G}$:
\begin{align}
 \mathbf{F}[\tilde{\Psi}](\mathbf{k}) &= (\mathbf{I}-\mathbf{\hat{k}}
 \mathbf{\hat{k}}^T) \int \mathbf{p'} (\mathbf{p'} \cdot
 \mathbf{\hat{k}}) \tilde{\Psi'} d\mathbf{p'} \\ \mathbf{G}
        [\tilde{\Psi}] (\mathbf{k}) &= \mathbf{\hat{k}} \int
        \tilde{\Psi'} d\mathbf{p'}.
\end{align}
$\mathbf{F}$ lies in the orthogonal subspace to $ \mathbf{\hat{k}}$
and hence to $\mathbf{G}$. Rewriting Eq. (\ref{PsiLinDecomp}) in
terms of these operators gives
\begin{align}\label{Psitilde_rt}
\tilde{\Psi}&= \frac{-3\alpha \gamma}{4 \pi} \frac{1}{\sigma + \lambda_0 
+ i \mathbf{p} \cdot \mathbf{k}} \mathbf{k}^T \mathbf{p} \mathbf{p}^T \mathbf{F}[\tilde{\Psi}] \nonumber \\
&+ \frac{\lambda_0}{4 \pi} \frac{\left( \frac{ik\chi \beta_2 }{\beta_1 + k^2 D_c} \mathbf{p}^T 
+ \mathbf{\hat{k}}^T  \right)}{\sigma + \lambda_0 + i \mathbf{p} \cdot \mathbf{k}}  \mathbf{G}[\tilde{\Psi}]
\end{align}
To obtain eigenvalue relations, apply $\mathbf{F}$ and $\mathbf{G}$ to $\tilde{\Psi}$:
\begin{align}
 \mathbf{F}[\tilde{\Psi}] &= -\frac{3 \alpha \gamma}{4 \pi}
 (\mathbf{I}-\mathbf{\hat{k}} \mathbf{\hat{k}}^T) \int \frac{
   (\mathbf{p'} \cdot \mathbf{\hat{k}})^2 \mathbf{p'} \mathbf{p'}^T
   d\mathbf{p'} }{\sigma + \lambda_0+ i (\mathbf{p'} \cdot
   \mathbf{k})} \mathbf{F}[\tilde{\Psi}] \nonumber \\ &+
 \frac{\lambda_0}{4 \pi} (\mathbf{I}-\mathbf{\hat{k}}
 \mathbf{\hat{k}}^T) \int \mathbf{p'} \frac{(\mathbf{p'} \cdot
   \mathbf{\hat{k}})}{\sigma + \lambda_0 + i (\mathbf{p'} \cdot
   \mathbf{k})} \nonumber \\ &\left( \frac{ik\chi \beta_2 }{\beta_1 +
   k^2 D_c} \mathbf{p'}^T + \mathbf{\hat{k}}^T \right) d\mathbf{p'}
 \mathbf{G}[\tilde{\Psi}]\label{FEqn_rt} 
 \end{align}
 and
 \begin{align}
 \mathbf{G} [\tilde{\Psi}] &= -\frac{3 \alpha \gamma}{4 \pi} \int
 \frac{ (\mathbf{p'} \cdot \mathbf{\hat{k}})^2 }{\sigma + \lambda_0
   +i\mathbf{p'} \cdot \mathbf{k}} d\mathbf{p'} \mathbf{F}
      [\tilde{\Psi}]\nonumber \\ &+\frac{\lambda_0}{4 \pi} \int \frac{
        \left( \frac{ik\chi \beta_2 }{\beta_1 + k^2 D_c} (\mathbf{p'}
        \cdot \mathbf{\hat{k}})+1 \right) }{\sigma +\lambda_0 +
        i\mathbf{p'} \cdot \mathbf{k}} d\mathbf{p'} \mathbf{G}
      [\tilde{\Psi}].\label{GEqn_rt}
\end{align}
These equations are invariant under rotations, so without loss of
generality we can choose a coordinate system in which
$\mathbf{\hat{k}}=\mathbf{\hat{z}}$. In spherical coordinates with
polar angle $\theta \in [0, \pi]$ and azimuthal angle $\phi \in
[0,2\pi)$, we have $\mathbf{p}=\left[ \sin \theta
    \cos \phi, \sin \theta \sin \phi, \cos \theta \right]$ and
  $d\mathbf{p} = \sin \theta d\theta d\phi$. Since $
  \mathbf{F}[\tilde{\Psi}]$ is perpendicular to $\mathbf{\hat{k}}$,
  Eq. (\ref{FEqn_rt}) becomes
\begin{align*}
\mathbf{F}[\tilde{\Psi}] = &-\frac{3 \alpha \gamma}{4 \pi}
\int_0^{\pi} \frac{\cos^2 \theta \sin^3 \theta}{\sigma + \lambda_0 +
  ik \cos \theta} d \theta \nonumber \\ &\int_0^{2\pi} \left( \cos^2
\phi \hat{\mathbf{x}}\hat{\mathbf{x}} + \sin^2 \phi
\hat{\mathbf{y}}\hat{\mathbf{y}} \right) d\phi
\mathbf{F}[\tilde{\Psi}].
\end{align*}
Since  $ \mathbf{G}(\tilde{\Psi})$ is in the $\mathbf{\hat{k}}$ direction, Eq. (\ref{GEqn_rt}) becomes
\begin{align*}
\mathbf{G}[\tilde{\Psi}] &= \frac{\lambda_0}{4 \pi}  \frac{ik\chi \beta_2 }{\beta_1 + k^2 D_c} \int_0^{\pi} \frac{\cos \theta \sin \theta d\theta}{\sigma + \lambda_0 + ik\cos\theta}  \int_0^{2\pi} d\phi   \mathbf{G}[\tilde{\Psi}]  \nonumber\\
&+ \frac{\lambda_0}{4 \pi}\int_0^{\pi} \frac{ \sin \theta}{\sigma + \lambda_0 + ik\cos\theta} d\theta \int_0^{2\pi} d\phi   \mathbf{G}[\tilde{\Psi}].
 \end{align*}

Performing the integrals in $\phi$ above, we obtain the two
\textit{separate} dispersion relations
\begin{align}
1 &= -\frac{3 \alpha \gamma}{4ik}\left[ 2a_H^3 -\frac{4}{3}a_H +
(a_H^4-a_H^2)\log \frac{a_H-1}{a_H+1} \right],\label{hydro-inst}\\
1 &=\frac{\lambda_0 \chi}{2}R\left[ 2+a_C\log \frac{a_C-1}{a_C+1} \right]
-\frac{\lambda_0}{2}\frac{1}{ik}\log \frac{a_C-1}{a_C+1}
\label{chem-inst}
\end{align}
where for simplicity we have defined $a=(\sigma + \lambda_0)/ik$ and
$R=\beta_2/(\beta_1+k^2 D_c)$. We refer to
Eqs. (\ref{hydro-inst}) \& (\ref{chem-inst}) as the hydrodynamic and
auto-chemotactic dispersion relations, respectively.

The relation in Eq. (\ref{hydro-inst}) is similar to that in
Saintillan \& Shelley \cite{SaintShelley08b} for non-chemotactic non-tumbling
swimmers, but with the addition of the basic tumbling frequency
$\lambda_0$. It has also been found independently by Subramanian
\& Koch for purely-tumbling swimmers \cite{SubKoch09}. The
auto-chemotactic relation in Eq.~(\ref{chem-inst}) is new
\cite{LGS12}. Note that chemotaxis enters the hydrodynamic relation
Eq.~(\ref{hydro-inst}) solely through stopping rate $\lambda_0$. The
fluid dynamics and its effects (e.g. the swimming mechanism typified
by the parameter $\alpha$) do not appear in the auto-chemotactic
relation Eq.~(\ref{chem-inst}), but the quasi-static chemo-attractant dynamics is included
in the term $ R= \beta_2/(\beta_1+k^2 D_c)$.

\subsubsection{Eigenmodes}
The previous analysis shows that $\mathbf{F}$, which is associated
with hydrodynamics, is perpendicular to $\mathbf{\hat{k}}$, while
$\mathbf{G}$, associated with chemotaxis, is parallel to it.  From
Eq.~(\ref{Psitilde_rt}) we can see that the eigenmodes of the
distribution function are linear combinations of the form
\begin{align}\label{eigenmode}
  \tilde{\Psi} = a_{1} \frac{(\mathbf{p} \cdot
    \mathbf{\hat{k}})(\mathbf{p} \cdot
    \mathbf{\hat{k}_{\perp}})}{\sigma +\lambda_0 + ik (\mathbf{p}
    \cdot \mathbf{\hat{k}})} + a_{2} \frac{ (\chi R~ik\mathbf{p} +
    \mathbf{\hat{k}}) \cdot \mathbf{\hat{k}} }{\sigma +\lambda_0 +
    ik(\mathbf{p} \cdot \mathbf{\hat{k}})},
\end{align}
where $\mathbf{\hat{k}_{\perp}}$ is any vector perpendicular to
$\mathbf{\hat{k}}$ and $a_{1}, a_{2}$ are constants for the
independent hydrodynamic and chemotactic contributions,
respectively. 

The above decomposition onto zeroeth and first azimuthal modes tells
us that concentration field fluctuations can arise from 
chemotactic processes. This is reflected through the fact that
\begin{equation}
 \tilde{\Phi}(\mathbf{k}, t) = \int \tilde{\Psi}(\mathbf{k},
 \mathbf{p},t) d \mathbf{p}= a_{2}(k) \int \frac{
   (\chi R~ik\mathbf{p} + \mathbf{\hat{k}}) \cdot \mathbf{\hat{k}} }{\sigma
   +\lambda_0 + ik(\mathbf{p} \cdot \mathbf{\hat{k}})} d \mathbf{p}
\end{equation}
is not generally zero. From simulations it is known that concentration
fluctuations in non-chemotactic suspensions can also develop
from the nonlinearities that linear analysis neglects
\cite{SaintShelley08b}.  If we substitute Eq.(\ref{eigenmode}) into
the definition of the active particle stress (the nondimensional form
of Eq.~(\ref{stress-nondim})) and make use of the dispersion relations
Eqs.~(\ref{hydro-inst},\ref{chem-inst}), we find that the active
stress eigenmodes are of the form
\begin{align}
\label{eigenmode_decomp}
 \tilde{\Sigma}^p = a_{1} \left( \mathbf{\hat{k}_{\perp}\hat{k}}^T 
+ \mathbf{\hat{k}\hat{k}_{\perp}}^T \right)   
+ a_{2} \left( \mathbf{\hat{k}\hat{k}}^T 
- \mathbf{\hat{k}_{\perp}\hat{k}_{\perp}}^T \right),
\end{align}
which is a sum of shear stresses (first term) and normal stresses
(second term). The normal stresses arise solely from chemotaxis,
while the shear stresses arise from hydrodynamics. Since $
\tilde{\mathbf{u}} = a_{1}(k) i/k \mathbf{\hat{k}} $, linear stability
analysis does not predict the growth of any velocity fluctuations due
to chemotaxis. That will happen due to nonlinearities, which is
examined through simulation.

\subsubsection{Long-wave asymptotic expansions}
To gain further insight into the system, we look at small $k$ (large system size)
asymptotic solutions for $\sigma(k)$ in the hydrodynamic and
chemotactic relations, Eqs. (\ref{hydro-inst}) and (\ref{chem-inst}),
respectively.
For the hydrodynamic relation Eq. (\ref{hydro-inst}) we obtain two
branches:
\begin{align}\label{hydro-asymp}
\sigma_{H1} \approx -\frac{\alpha\gamma}{5}-\lambda_0 +\frac{15}{7\alpha \gamma} k^2 + O(k^3) \nonumber \\
\sigma_{H2} \approx  -\lambda_0- \frac{1}{\alpha \gamma}k^2 +O(-\alpha k^3).
\end{align}
The auto-chemotactic relation Eq. (\ref{chem-inst}) gives only one
branch that satisfies the integral relation Eq. (\ref{GEqn_rt}):
\begin{align}\label{chem-asymp}
\sigma_C &\approx \frac{1}{3\lambda_0} (\overline{\chi}\lambda_0 -1 )k^2 \nonumber \\
&- \frac{(5 \overline{\chi}^2 \lambda_0^2-6\overline{\chi}\lambda_0 +1 + 15\overline{\chi} \overline{D_c} \lambda_0^3 )}{45 \lambda_0^3}k^4 + O(k^5),
\end{align}
with $\overline{\chi} = \chi \beta_2/\beta_1$ and
$\overline{D_C}=D_C\beta_2/\beta_1$.

From Eq. (\ref{hydro-asymp}) we can infer that there is an instability
at large system sizes arising from the hydrodynamics, but not from
chemotaxis.  Nonetheless, from Eq. (\ref{chem-asymp}), we can obtain a
range of parameters for which $\sigma_C>0$ for $k>0$ and so find a
range of parameters that give a
chemotactic instability. This occurs for $\overline{\chi}\lambda_0 >
1$, or more specifically for $\chi \beta_2 \lambda_0/\beta_1
>1$. Chemo-attractant diffusion comes in at the next-order term in $k$ in
Eq. (\ref{chem-asymp}) and has a stabilizing effect.

\subsubsection{Solving the Dispersion Relation}

We solve the dispersion relations in
Eqs. (\ref{hydro-inst}, \ref{chem-inst}) for $\sigma(k)$ numerically using
Newton's method. An eigenvalue estimate found at small $k$ using
the asymptotic solutions of Eqs. (\ref{hydro-asymp}) and
(\ref{chem-asymp}) is used as an initial guesses. 
We then march to larger $k$, where at
each Newton solve the iteration is started with the converged solution
at the previous, smaller value of $k$.  The numerical solutions are
checked to make sure that they satisfy the integral relations
Eq. (\ref{FEqn_rt}) and Eq. (\ref{GEqn_rt}).

\begin{figure}[htps]
\centering
\includegraphics[width=3.3in, height = 2.75in]{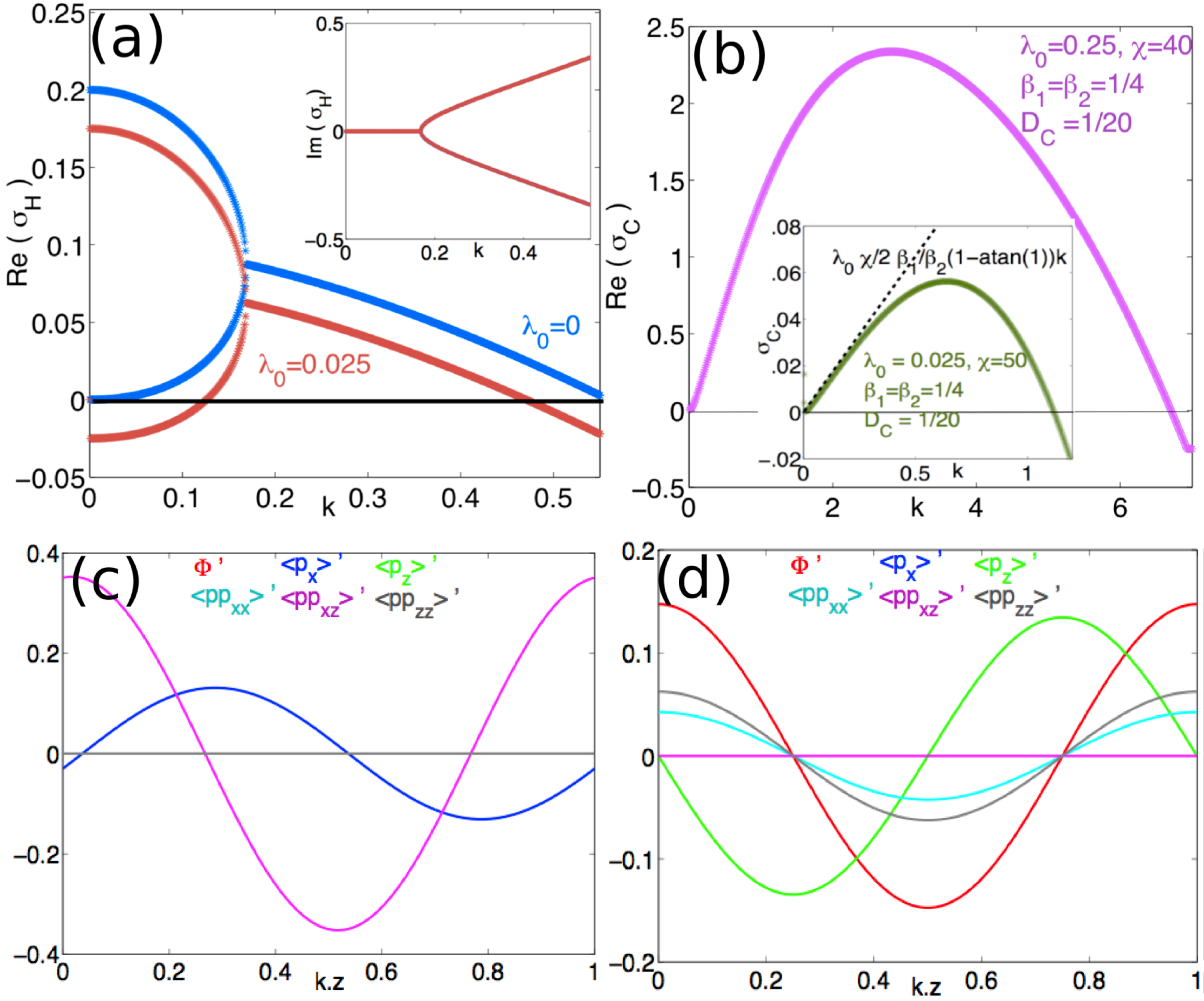}
\caption{ (color online) (a) The two branches of the growth rate
  obtained by the hydrodynamic relation. Inset shows the imaginary
  parts. (b) The growth rate obtained from the chemotaxis relation for
  $\lambda_0=0.25, \chi=40, \beta_1=\beta_2=1/4, D_c=1/20$. Inset
  shows the chemotaxis branch for parameters $\lambda_0=0.025,
  \chi=50, \beta_1=\beta_2=1/4, D_c=1/20$ used for simulations. 
  The dashed line in the inset and its denoted slope are for a later 
  comparison in Section IIID with the TP model. 
  (c) Unstable eigenmodes or perturbations in the zeroeth, first 
  and second moments of the distribution function w.r.t. the 
  orientation vector $\mathbf{p} $ due to the
  hydrodynamics instability for wavenumber $\mathbf{k} = 0.2\mathbf{\hat{k}}$ in
  a suspension of pushers with $\alpha = -1, \gamma = 1$, and basal
  tumbling frequency $\lambda_0= 0.025$. See Section IIIA-5 for 
  an explanation. (d) Unstable 
  eigenmodes or perturbations in the zeroth, first 
  and second moments of the distribution function w.r.t. the 
  orientation vector $\mathbf{p} $  due to the auto-chemotactic
  instability for wavenumber $\mathbf{k} = 0.2\mathbf{\hat{k}}$ a suspension of
  swimmers with $\gamma = 1$, $\lambda_0=
  0.025$, $\chi=50, \beta_1=\beta_2=1/4,
  D_c=1/20$. 
}\label{fig:GrowthRates}
\end{figure}

The solution to the hydrodynamics eigenvalue relation is shown in
Fig.~\ref{fig:GrowthRates}a for rod-like non-tumbling swimmers ($\gamma=1$
and $\lambda_0=0$), and tumbling swimmers with basic stopping rate
$\lambda_0=0.025$. The branch $\lambda_0=0$ is exactly that obtained
earlier by \cite{SaintShelley08b} for non-tumbling
swimmers.  The addition of tumbling merely shifts down the two
branches of $Re(\sigma_H)$ by $\lambda_0$, hence tumbling by itself
has a stabilizing effect. Tumbling does not affect the oscillatory
modes, as seen by the unaltered branches of $Im(\sigma_H)$.

For pushers ($\alpha<0$) there is a hydrodynamic instability for a
finite band of wavenumbers $k\in[0,k_c \approx 0.55)$. Tumbling
diminishes this range of unstable wave-numbers as the branch is
brought down by $\lambda_0$ (again $Im(\sigma_H)$ remains
unaffected). From the numerical solution, we can estimate the critical
wavenumber $k_c$ (and hence the critical system size $L_c=2\pi/k_c$)
for which $Re(\sigma(k_c))=0$.  Moreover, we can obtain a range of
$\lambda_0$ for which a hydrodynamic instability is possible for
pushers. We find that for $\lambda_0 \geq 0.2$ there can be no
hydrodynamic instability for any system size or swimmer shape (as
represented by $\gamma$).

There are two positive $Re(\sigma_H)$ branches which merge at
$k\approx 0.19$, and at which the two conjugate branches of
$Im(\sigma_H)$ bifurcate from zero.  This tells us that for $k<0.19$
we have two standing and growing modes, whereas beyond this the modes
become complex.

For pullers, there is no hydrodynamic instability, as
$Re(\sigma_H(k))<0$ even for $\lambda_0=0$ (see also
\cite{SaintShelley08b}). The addition of basal tumbling $\lambda_0$
again shifts down the $Re(\sigma_H)$ branches by $\lambda_0$ for all
wavenumbers $k$ and further stabilizes the system. $Im(\sigma_H)$
remains unchanged from the pusher case shown in
Fig.~\ref{fig:GrowthRates}a.

For the chemotactic dispersion relation, the long-wave asymptotics in
Eq.~(\ref{chem-inst}) yields growth rate $\sigma_C \approx
1/(3\lambda_0) [(\chi \beta_2/\beta_1) \lambda_0 -1 ]k^2 + O(k^4)$,
which shows that for $(\chi \beta_2/\beta_1) \lambda_0 >1$ there are
wavenumbers $k$ with $Re(\sigma_C(k))>0$, for pushers and pullers
alike and for any swimmer shape parameter $\gamma$. Auto-chemotaxis
thus introduces a new instability branch, which is solved numerically
from Eq.(\ref{chem-inst}) and plotted in Fig.~\ref{fig:GrowthRates}b
for two sets of $\lambda_0, \chi, D_c$.

\subsubsection{The Linear Perturbations}

Using Eq.~(\ref{eigenmode_decomp}), and the numerical solution 
of the dispersion relations Eqs.~(\ref{hydro-inst}, \ref{chem-inst}) 
for the parameters used in Fig.~\ref{fig:GrowthRates}a~\&~b, we 
have calculated the eigenmode using $\mathbf{\hat{k}}=\mathbf{\hat{z}}$ and
$\mathbf{\hat{k}}_{\perp}=\mathbf{\hat{x}}$. We consider the hydrodynamic and
auto-chemotactic contributions separately for $k=|\mathbf{k}|=0.2$.  
Fig.~\ref{fig:GrowthRates}c \& d illustrates the perturbations in the 
swimmer concentration $\Phi'$ as well as	 the components of the first moment 
vector (also the un-normalized polarity vector) and second moment tensor (proportional to the active stress -- neglecting the diagonal contribution) of the distribution function $\Psi$, namely
$<\mathbf{p}>$  and $<\mathbf{pp}>$  for values of $kz = \mathbf{k}\cdot \mathbf{x} \in [0,1]$.

As seen in Fig.~\ref{fig:GrowthRates}c, the hydrodynamic instability gives rise to
perturbations in the director field $<\mathbf{p}>'_x$ and
the off-diagonal elements of the second moment of the distribution
function w.r.t. the orientation vector, $<\mathbf{pp}>'_{xz}$. The
second moment is related to the active stress, as seen in
Eq.~(\ref{stress-nondim}), or linearized in Eq.~(\ref{stress_lin}),
thus perturbations in the active stress do arise for non-zero
$\alpha$. From Eq. (\ref{vel_lin}) we expect nonzero perturbations in
the fluid velocity to arise.  (See also \cite{SaintShelley08b}.) Perturbations 
due to the hydrodynamic instability are zero for the the other quantities, 
namely $\Phi'$ $<\mathbf{p}>'_z$, $<\mathbf{pp}>'_{xx}$ and
 $<\mathbf{pp}>'_{zz}$.

From Fig.~\ref{fig:GrowthRates}d we see that the chemotactic 
instability gives rise to perturbations in the zeroeth moment of the
 distribution function (which is the swimmer concentration), the
director field $<\mathbf{p}>'_z$ and the diagonal elements of the
second moment of the distribution function w.r.t. the orientation
vector, $<\mathbf{pp}>'_{xx}$ and $<\mathbf{pp}>'_{zz}$, which are
related to the normal stresses. Perturbations due to the chemotactic 
instability are zero for the other quantities, such as $<\mathbf{p}>'_x$ 
and $<\mathbf{pp}>'_{xz}$.

In the linearized system the hydrodynamic and chemotactic 
instabilities are uncoupled and operate independently in different directions. 
The hydrodynamic instability gives rise to shear stresses and a flux in
the $\mathbf{\hat{k}}_{\perp}$ direction. Auto-chemotaxis gives rise
to aggregation, a flux in the $\mathbf{\hat{k}}$ direction, and normal stresses. 
In the full non-linear system these perturbations and modes are of course coupled
and interesting dynamics emerges from the interplay of these instabilities, as shown 
later with simulations.

\subsection{A Phase Diagram of the Dynamics}

\begin{figure}[htps]
\vspace{-.1in}
 \centering
\includegraphics[width=3.3in, height=1.5in]{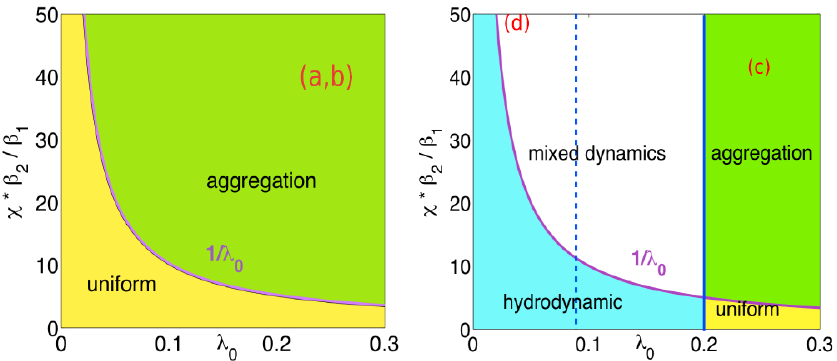}
\vspace{-.1in}
 \caption{ (Color online) Phase space of various regimes for
   auto-chemotactic and/or hydrodynamic instabilities in suspensions of
    pullers or neutral swimmers (left), and pushers (right) as a function of the basal tumbling frequency $\lambda_0$ and chemotactic parameters $\chi, \beta_1, \beta_2$. Solid boundaries refer to the
   linear stability at long waves. Dashed lines show where we observe
   the boundaries to shift in simulations at finite box size. Red 
   marks (a-d) indicate the parameters used for nonlinear simulations. }
   \vspace{-.1in}
 \label{fig:phasespace}
 \end{figure}

Linear theory shows that there is a range of $\lambda_0$ for
which there is a hydrodynamic instability in pusher suspensions. If $\lambda_0
\geq 0.2$, there is no hydrodynamic instability for any system size
and any swimmer shape $\gamma$, since, as seen in
Fig.~\ref{fig:GrowthRates}a, $Re(\sigma_H(k))\leq 0.2$. For an
auto-chemotactic instability we need $(\chi \beta_2/\beta_1) \lambda_0
>1$. This connects the auto-chemotaxis parameters $\chi$, $\beta_1$,
$\beta_2$ to the basal tumbling rate $\lambda_0$. This information
about the parameters is assembled in a phase diagram in Fig.~\ref{fig:phasespace}, which
shows the parameters and various dynamical regimes we expect
based on the linear theory and simulations.
 
 For pullers or neutral swimmer (right panel of
 Fig.~\ref{fig:phasespace}) there are only two states: aggregation due
 to auto-chemotaxis and the uniform stable state.  For pushers (right
 panel of Fig.~\ref{fig:phasespace}), there are four dynamical
 regimes: aggregation (due to auto-chemotaxis), a regime where strong
 mixing fluid flows emerge (due to the hydrodynamic instability), a
 mixed state of both auto-chemotactic and hydrodynamic instabilities,
 and the uniform stable state where perturbations should decay to
 uniform isotropy.  Other parameters, say the shape parameter
 $\gamma$, can alter this diagram.

\subsection{Linear Stability of Turning-Particle Auto-Chemotaxis}
\begin{figure}[htps]
\centering
\includegraphics[width=3.3in,height=1.5in]{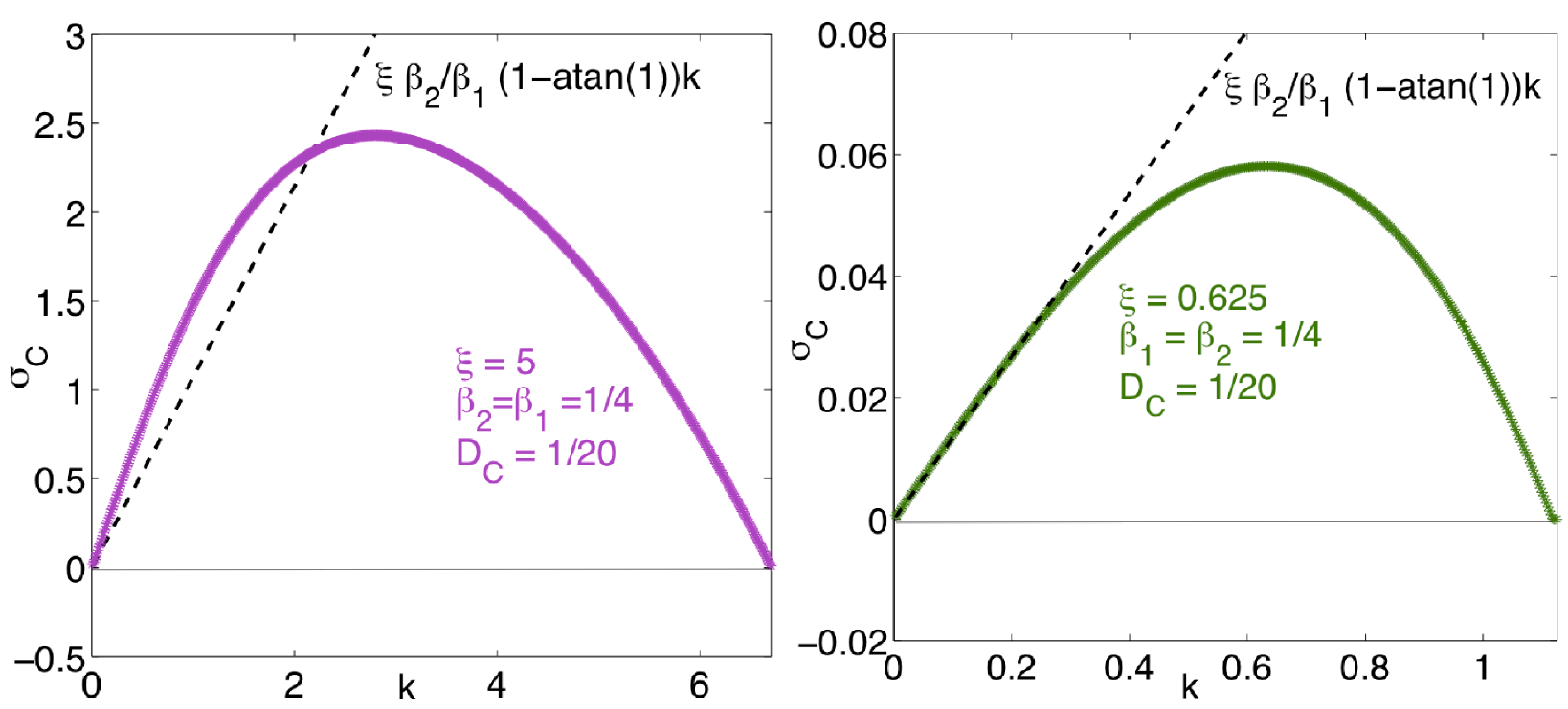}
   \caption{ (color online) The growth rate for the Turning-Particle 
   auto-chemotaxis relation: $\xi=5$ (left) and $\xi=0.625$ (right) 
   with $\beta_1=\beta_2=1/4, D_c=1/20$. Dashed lines show the 
   long-wave asymptotics. Parameters $\xi$ are chosen so that
   $\xi = \lambda_0 \chi/2$ with $\lambda_0,\chi$ corresponding 
   to those shown in Fig.\ref{fig:GrowthRates}b. 
   The comparison is discussed in Section IIID.}
   \label{fig:TPGrowthRates} 
 \end{figure}

The linear stability analysis of the Turning-Particle model of
auto-chemotaxis is done in a similar manner, and the results obtained are
remarkably similar to the Run-and-Tumble model with linearized
tumbling rate, even though there is tumbling in this instance. 
For swimmers with no translational or rotational diffusion ($D= d_r=0$), 
the two dispersion relations are
\begin{align}
 1 &= -\frac{3 \alpha \gamma}{4ik}\left[ 2a_H^3 -\frac{4}{3}a_H +
(a_H^4-a_H^2)\log \frac{a_H-1}{a_H+1} \right], \label{firstDispRel}\\
  1 &= \xi R \left[ 2+
    a_C\log\frac{a_C-1}{a_C+1} \right],\label{TP-chem-inst}
\end{align}
with $a= -i \sigma /k$. The long-wave (small $k$) asymptotics for the
hydrodynamics relation yields $\sigma_{H1} =-\frac{\alpha\gamma}{5}
+\frac{15}{7\alpha \gamma} k^2 +...$ and $\sigma_{H2} =-
\frac{1}{\alpha \gamma}k^2 +...$ which is similar to the results found
for the Run-and-Tumble model in Eqs. (\ref{hydro-asymp}) without basal
tumbling. The asymptotics of the auto-chemotactic relation
gives $\sigma_C = \sigma_1 k + \sigma_3 k^3+...$ with $\sigma_1
\approx \xi (1-\arctan(1) )\beta_2/\beta_1$ and $\sigma_3 \approx
D_c/\beta_1$. While this does not look similar to the Run-and-Tumble
result in Eq. (\ref{chem-asymp}), the numerical solution in Fig.~\ref{fig:TPGrowthRates}
shows similarities in the overall curve shape, maxima and critical
$k_c$ where $\sigma (k_c)=0$.

Moreover, a finite band of wavenumbers with $Re(\sigma_C)>0$ can be
found for $k\in[0,k_c=\sqrt{(2\xi\beta_2-\beta_1)/D_c)}]$. For
this model we can as well express the eigenmodes for the distribution
function as linear combinations of the form in
Eq. (\ref{eigenmode}) with the eigenvector arising from chemotaxis in
the $\mathbf{\hat{k}}$ direction and the eigenvector arising from the
hydrodynamics perpendicular to $\mathbf{\hat{k}}$. It can also be
shown that the hydrodynamic instability increases growth of the
shear stresses whereas the auto-chemotactic instability increases
fluctuations in the swimmer concentration and normal stresses.

 Although not shown here, including translational diffusion with
 constant $D$ merely shifts down the $Re(\sigma_H)$ and $Re(\sigma_C)$
 by $-Dk^2$ (see \cite{SaintShelley08b} for non-tumbling
 non-chemotactic swimmers). As found by \cite{HohenShelley10}, when
 rotational diffusion with constant $d_r$ is included in the stability
 analysis of non-chemotactic swimmers, the hydrodynamic instability
 branch $Re(\sigma_H)$ shifts down by approximately $6d_r$ for small
 $k$.  We do not discuss here how non-zero rotational diffusion $d_r$
 affects the chemotactic instability. 

\subsection{Relating the Two Chemotaxis Models}

In the Run-and-Tumble chemotaxis model, tumbling helps stabilize the
the system; the hydrodynamic instability branches are shifted
downwards by the basal tumbling frequency $\lambda_0$, as seen in
Fig. \ref{fig:GrowthRates}.  And, as just mentioned above, rotational
diffusion shifts down the hydrodynamic instability branches in
non-chemotactic suspensions by approximately $6d_r$
\cite{HohenShelley10, LushiDissert} for $k\ll1$. In this respect, at
large system sizes tumbling with basal frequency $\lambda_0$ acts like
rotational diffusion with coefficient $6d_r$.

Comparing the first terms of the two chemotactic dispersion relations,
Eqs. (\ref{chem-inst}) \& (\ref{TP-chem-inst}), suggests that $\xi
\approx \lambda_0 \chi/2$ relates the behavior of the RT model with
basal tumbling $\lambda_0$ and chemotactic strength $\xi$ to the
behavior of the TP chemotaxis model with strength $\xi$. Since in the
$k\ll 1$ regime the chemotactic growth rate of the TP model is
$\sigma_C \approx \xi (1-\arctan(1))\beta_2/\beta_1k$, we plot the
line with slope $\lambda_0 \chi/2 (1-\arctan(1))\beta_2/\beta_1$ in
Fig.~\ref{fig:GrowthRates}b and see that it gives a good approximation
to the growth rate from the RT model in the $k\ll 1$ limit.
Comparison of the curves in Fig.~\ref{fig:GrowthRates}b for the RT
model and Fig.~\ref{fig:TPGrowthRates} for the TP model, when the
chemotactic parameters are matched as such, shows also their
similarity in overall curve shape, maxima and critical wave-numbers
$k_c$ where $\sigma (k_c)=0$. The eigenmodes of the two models
also have a very similar structure.

Thus, for $k\ll1$, the linearized TP model with chemotactic parameter
$\xi$ and rotational diffusion $d_r$ behaves similarly to the
linearized RT model with basal tumbling $\lambda_0$ and chemotactic
sensitivity $\chi$, when the parameters are related as $\xi \approx
\lambda_0 \chi/2$ and $\lambda_0=6d_r$. Full nonlinear simulations for
box sizes corresponding to $k\ll1$ and with parameters chosen as above
support this matching, as is shown in the next section.

\section{Nonlinear Simulations}
\label{Simulations}
In full simulations we focus primarily on the Run-and-Tumble model,
leaving the Turning-Particle model for comparison at the end,
but noting we expect similar dynamics when the parameters
 are properly matched as explained in Section IIID.
Using simulation we investigate the system dynamics in the various
regimes suggested by phase diagram in Fig.~\ref{fig:phasespace}.  Of particular interest
is the {\em aggregation} regime for different types of swimmers, and
the {\em mixed dynamics} regime of pusher suspensions.

\subsection{Numerical Method}

For relative ease of simulation we consider 2D doubly periodic systems
in which the particles are constrained to the $(x,y)$-plane with
orientation parametrized by an angle $\theta \in [0, 2\pi)$ so
$\mathbf{p} = (\cos \theta, \sin \theta, 0)$.  Since all the dependent
variables are periodic in $x$, $y$ and $\theta$ we use discrete
Fourier transforms (via the FFT algorithm) to approximate spatial and
angular derivatives and to solve the flow equations
Eq.~(\ref{Stokes-dim}). Integrations in $\theta$ to obtain the
swimmer density $\Phi$ (\ref{PhiEqn}) and active particle stresses
$\Sigma^a$ [Eq.~(\ref{stress-nondim})] use the trapezoidal rule, which
is spectrally accurate in this instance. Usually $128-256$ points are
used in the $x$ and $y$ directions and $32-64$ in the $\theta$
direction. We integrate the distribution equation
Eq.~(\ref{runandtumble3D}) and the chemo-attractant
advection-diffusion equation Eq.~(\ref{chemo}) using a second-order
time-stepping scheme
(Adams-Bashforth/Crank-Nicholson). Particle translational and
rotational diffusion and chemo-attractant diffusion are included in
all simulations for numerical stability (values of
$D=d_r=0.025$ and $D_c=0.05$ were used in most simulations). All
results presented are for slender micro-organisms with $\gamma=1$ and
the spatial square box side is $L=50$. The 
initial swimmer distribution is taken to be a uniform and isotropic
suspension perturbed as
\begin{align}\label{Psi_init}
 \Psi(\mathbf{x},\theta,0) = \frac{1}{2 \pi} \left[ 1+ \sum_i \epsilon_i \cos (\mathbf{k}_i \cdot \mathbf{x} + \xi_i)P_i(\theta) \right],
\end{align}
where $\epsilon_i$ is a small random coefficient
($\epsilon_i\in[-0.01,0.01]$), $\xi_i$ is a random phase and
$P_i(\theta)$ is a third-order polynomial of $\sin \theta$ and $\cos
\theta$ with random $O(1)$ coefficients. The initial chemo-attractant
concentration is taken to be uniform and $C(\mathbf{x},0) = \beta_2 /
\beta_1$ with $\beta_1=\beta_2$ used for most simulations.

\newpage

\subsection{Chemotaxis-Driven Dynamics}

\begin{figure*}[htps]
\vspace{-0.2in}
\centering
\includegraphics[width=6in, height=3.5in]{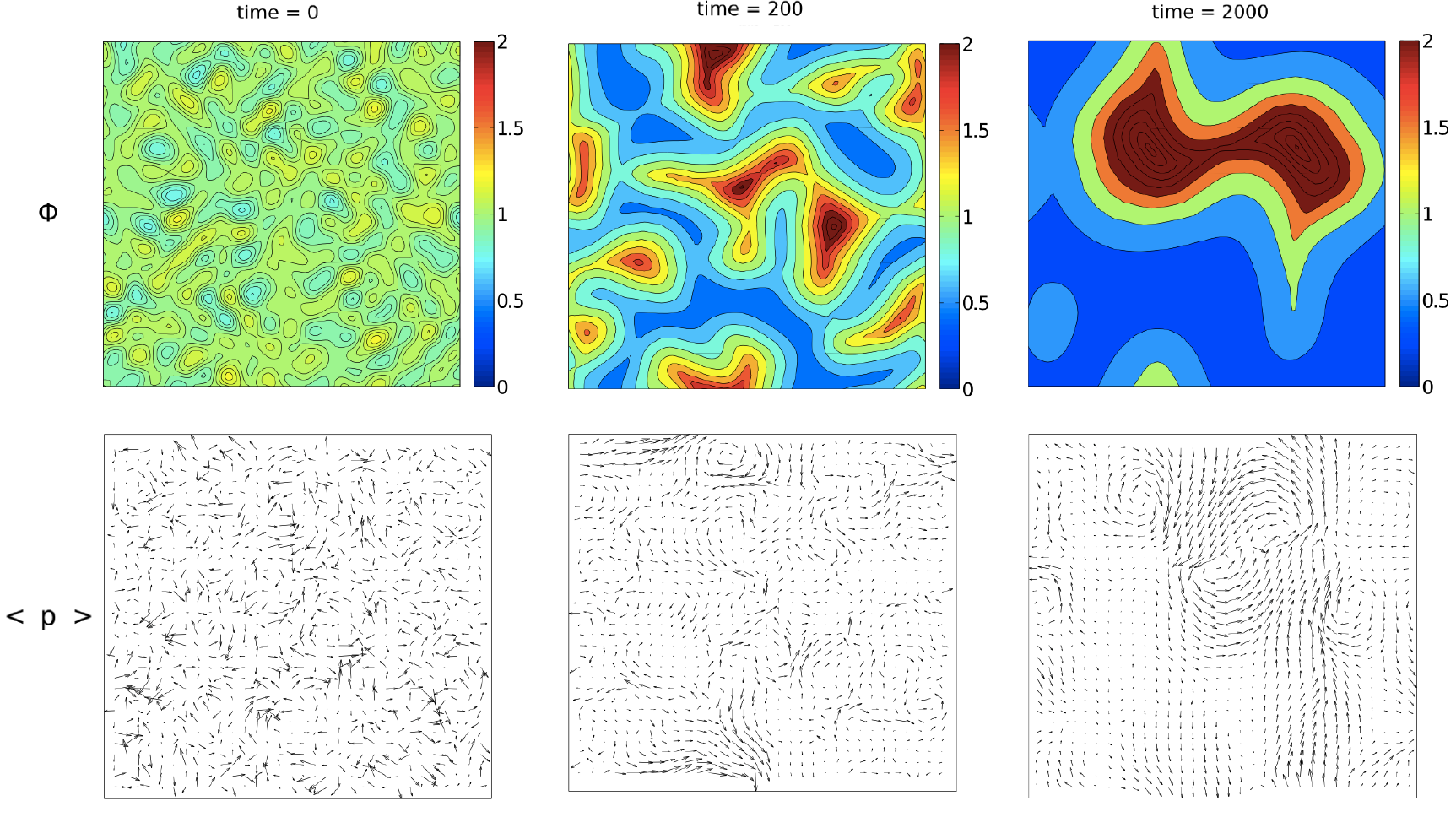}
\vspace{-0.15in}
  \caption{ (Color online) Auto-chemotaxis in a neutral swimmer
    suspension at times $t=$0, 200, and 2000. The first row shows
    concentration of swimmers $\Phi$, the second row shows the swimmer
    mean direction $<\mathbf{p}>$. At $t=$200, $max|<\mathbf{p}>|=0.091$
     and at $t=$2000, $max|<\mathbf{p}>|=0.086$.}
\vspace{-0.1in}
 \label{fig:auto_neut}
\end{figure*}

\begin{figure*}[htps]
\vspace{-0.2in}
\centering
\includegraphics[width=5in, height =1.6in]{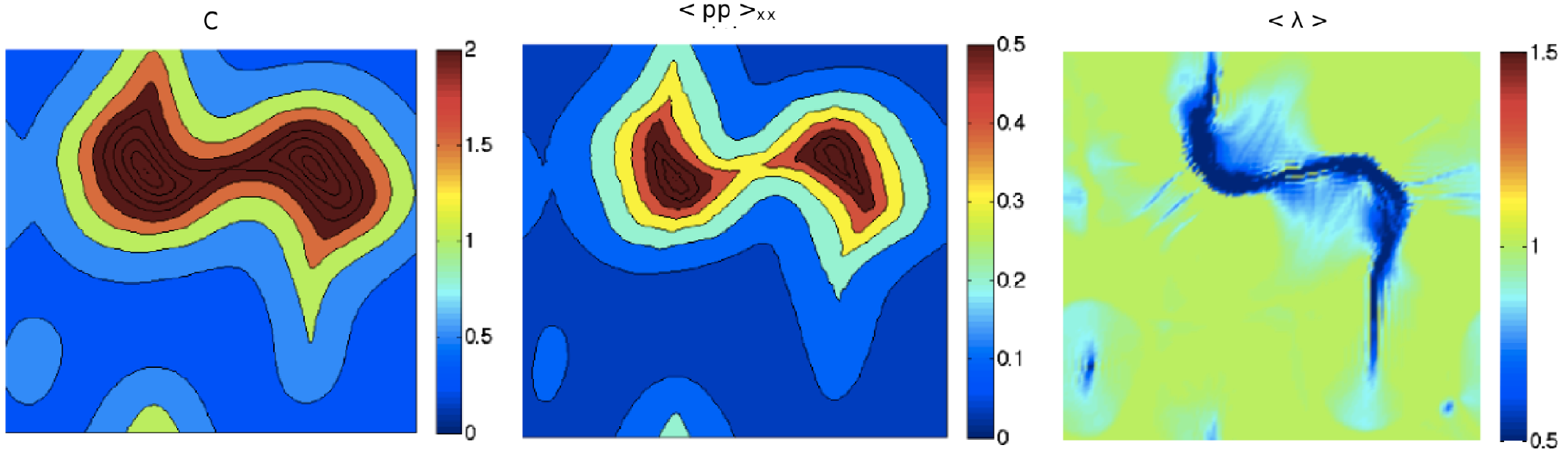}
 \vspace{-0.15in}
  \caption{ (Color online) Auto-chemotaxis in a neutral swimmer
    suspension at time $t=2000$. Shown are the chemoattractant
    concentration $C$, the diagonal entry $<\mathbf{pp}>_{xx}$
    of the second moment tensor of the distribution function,
    and the mean tumbling frequency $<\lambda(\mathcal{D}_t C)>$.}
\vspace{-0.1in}
 \label{fig:auto_neut1}
\end{figure*}

To illustrate the effect of pure auto-chemotaxis in the absence of
hydrodynamics, we first simulate a suspension of theoretical neutral swimmers by
setting $\alpha=0$. The other parameters used are $\lambda_0=0.25$,
$\chi=40$, $\beta_1=\beta_2=1/4$ for which the linear stability
analysis suggests an aggregation
instability -- see the red mark (a) in the phase diagram of 
Fig.~\ref{fig:phasespace}. Fig.~\ref{fig:auto_neut} shows plots of the swimmer
concentration $\Phi$ and mean direction $<\mathbf{p}>=\int
d\mathbf{p'} \Psi(\mathbf{p'})\mathbf{p'}/\Phi$ at 
times $t=$0, 200, and 2000. The chemo-attractant
concentration field $C$, the diagonal entry $<\mathbf{pp}>_{xx}$ 
of the second moment tensor  
(related to the active stress, were it present in this case), and the mean tumbling rate
$<\lambda(\mathcal{D}_t C)>$, are shown at $t=2000$ in
Fig.~\ref{fig:auto_neut1}.

The swimmer concentration field in Fig. \ref{fig:auto_neut}
shows continual aggregation and coarsening. At late times
($t=2000$) the swimmers have aggregated into a single still-evolving
doubly-peaked irregular mass. The chemo-attractant field closely
follows the swimmer concentration as seen in
Fig.~\ref{fig:auto_neut1}, with the two plots being nearly identical
to the eye. Indeed, this observation supports the validity of the approximation of 
a quasi-static chemo-attractant field in the linear stability
analysis. Note that $<\mathbf{pp}>_{xx}$ is also
similar to the swimmer concentration field, being significant only in
the aggregation regions, suggesting a link between hydrodynamic
interactions (were they present) and chemotactic aggregation. Within
the aggregation region the chemo-attractant gradient is steep and the
mean tumbling frequency $<\lambda(\mathcal{D}_t C)>$ is less
than the basal frequency $\lambda_0$, suggesting further swimmer
aggregation and coarsening.

While continual aggregation is observed, there is little sign of the
rapid self-focussing associated with the finite-time chemotactic collapse
\cite{Childress84, JagerLuck92,DolPert04} seen in the KS model. Here
that may in part be due to the constant swimming speed of individual
particles \cite{SchnitzerEtAl90}. 

\subsection{Auto-chemotaxis of Puller Suspensions}

\begin{figure*}[htps]
\vspace{-0.2in}
\centering
   \includegraphics[width=6in, height=3.5in]{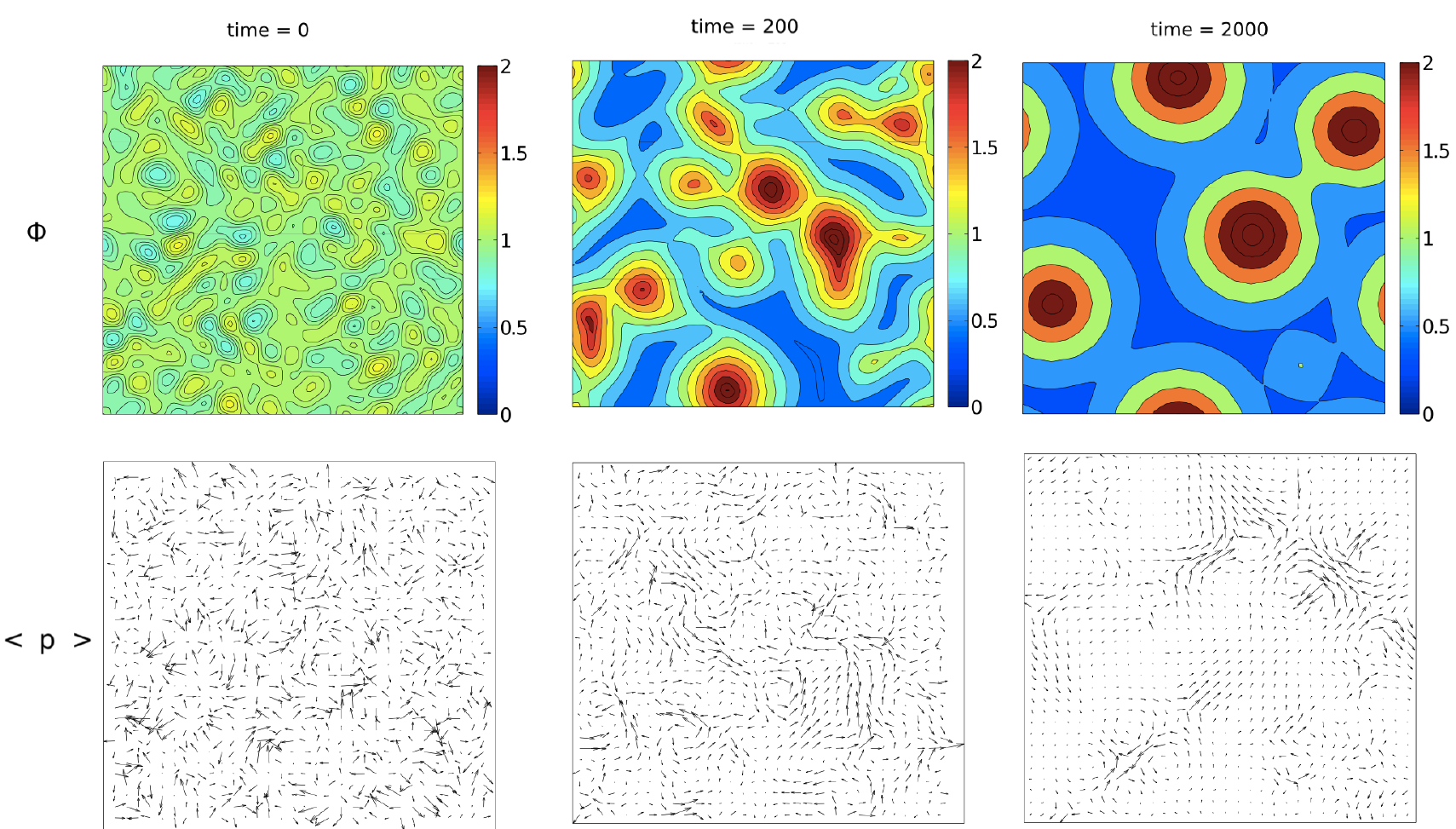}
   \vspace{-0.15in}
  \caption{ (Color online) Auto-chemotaxis in a puller swimmer
    suspension at times $t=0,200,2000$. The first row shows
    concentration of swimmers $\Phi$, the second row shows the swimmer
    mean direction $<\mathbf{p}>$. At $t=$200, $max|<\mathbf{p}>|=0.085$ and at $t=$2000, $max|<\mathbf{p}>|=0.067$.}
    \vspace{-0.1in}
 \label{fig:auto_pull}
\end{figure*}

\begin{figure*}[htps]
\vspace{-0.2in}
\centering
  \includegraphics[width=5in, height =1.6in]{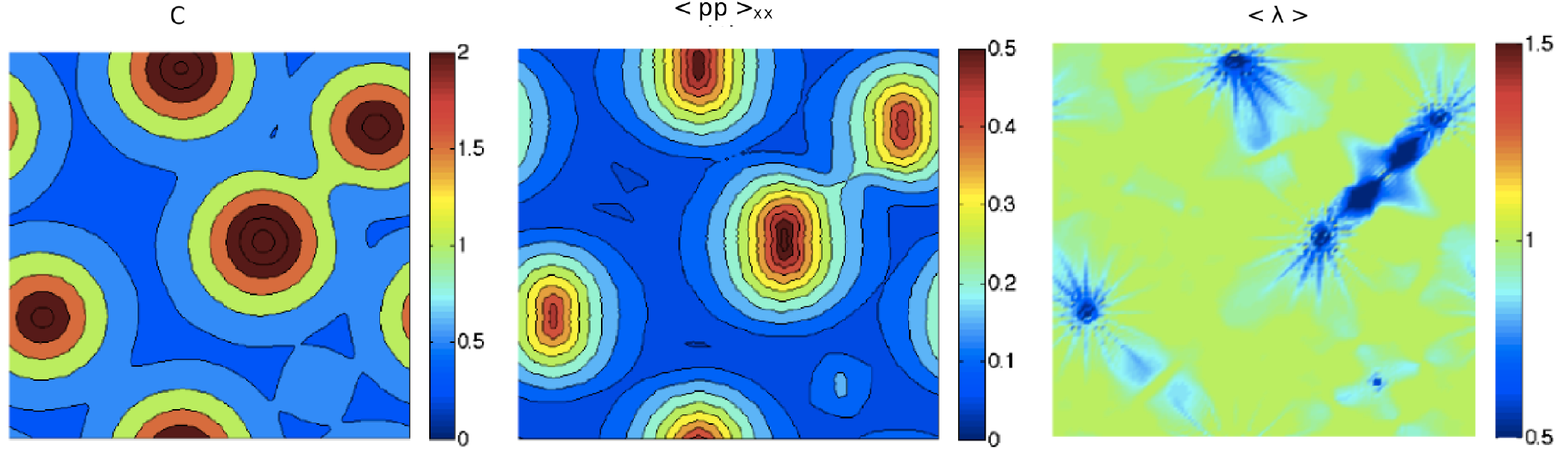}
  \vspace{-0.15in}
\caption{ (Color online) Auto-chemotaxis in a puller swimmer suspension at time $t=2000$. Shown are the chemoattractant concentration $C$, the diagonal entry $<\mathbf{pp}>_{xx}$
    of the second moment tensor of the distribution function, and the mean tumbling frequency $<\lambda(\mathcal{D}_t C)>$. } 
    \vspace{-0.1in}
 \label{fig:auto_pull1}
\end{figure*}

\begin{figure}[htps]
\centering
  \includegraphics[width=3.2in, height=3.0in]{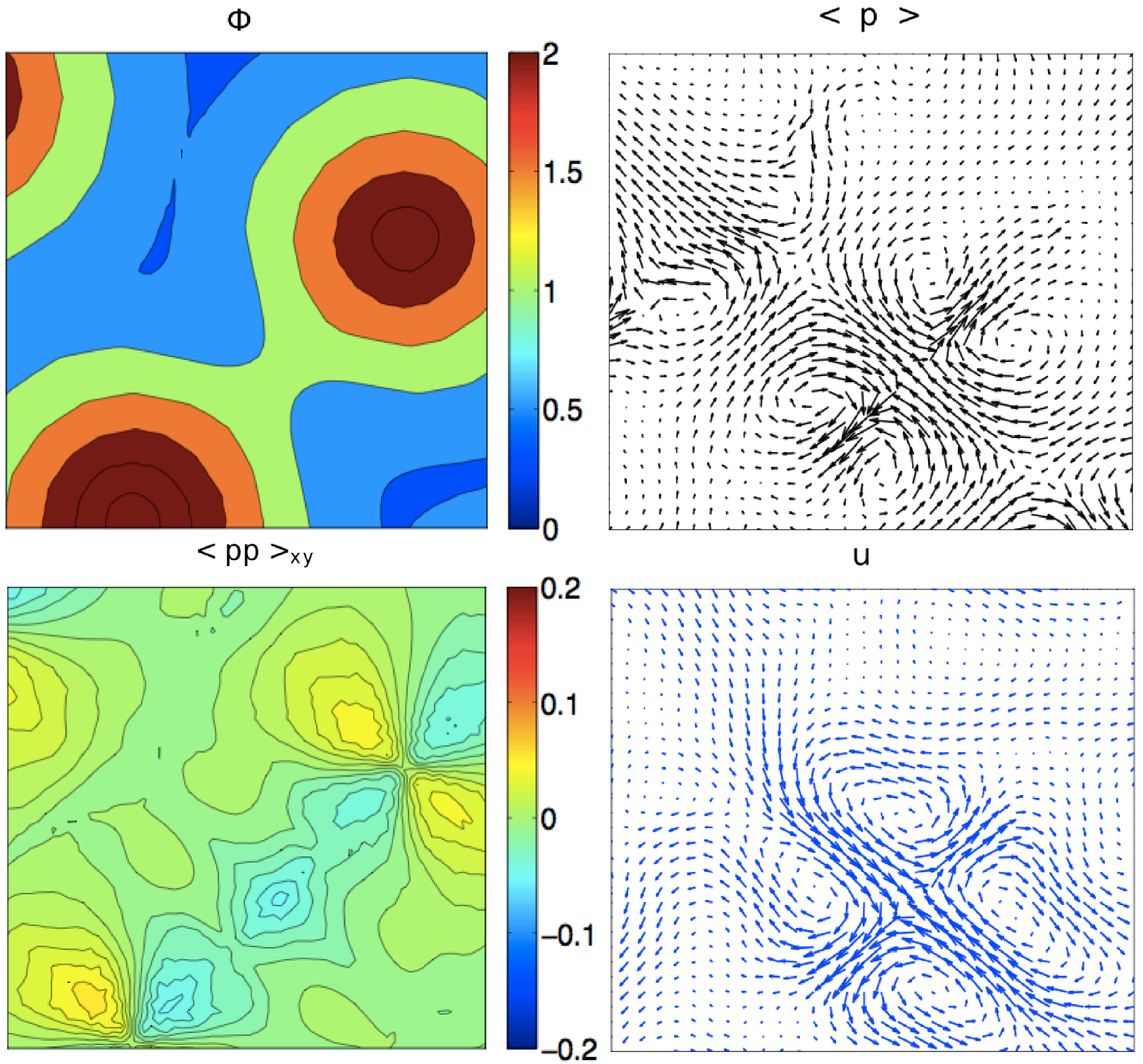}
  \vspace{-0.1in}
  \caption{ (Color online) Closeup of the top-right quarter of the
    domain of the auto-chemotactic puller suspension dynamics at time
    $t=2000$ as shown in Fig. \ref{fig:auto_pull}. Top: the swimmer concentration $\Phi$
    and mean direction $<\mathbf{p}>$ with $max|<\mathbf{p}>|=0.067$. Bottom: the off-diagonal entry of the second moment tensor $<\mathbf{pp}>_{xy}$  (related to the active stress) and the generated fluid velocity
    $\mathbf{u}$ with $max|\mathbf{u}|=0.03$.}
    \vspace{-0.1in}
 \label{fig:auto_pull3}
\end{figure}

We next perform the same numerical experiment but with a suspension of
pullers (setting $\alpha=1$). The other system parameters remain
unchanged -- see the red mark (b) in the phase diagram 
of Fig.~\ref{fig:phasespace}. Linear stability analysis suggests an aggregation
instability due to auto-chemotaxis, but with hydrodynamical
interactions suppressed since the growth rates in the hydrodynamic
instability have negative real part.  The results are illustrated in
Figs.~\ref{fig:auto_pull} and \ref{fig:auto_pull1}.

Fig. \ref{fig:auto_pull} for time $t=200$ shows that initial aggregation and
coarsening occurs in this suspension of pullers in a way very
similar to that of neutral swimmers. The swimmers aggregate into a few
regions that merge and grow. As before, the chemo-attractant
concentration field closely follows swimmer concentration, and 
$<\mathbf{pp}>_{xx}$ (related to the
active first normal stress) is significant in the aggregation
regions. There is a chemo-attractant spatial gradient in the
aggregation regions at early times and the swimmer mean tumbling
frequency there is less than the basal frequency $\lambda_0$,
suggesting the swimmers are on average moving toward higher values of
chemoattractant concentration.

However major differences from the neutral swimmer case eventually
emerge in the pattern morphology. At late times ($t=2000$) the
aggregation regions have become circular and there is no indication of
further coarsening. Indeed, these circular regions are all of a
similar size and appear to be mutually repelling, suggesting that this
is the terminal state of the system. Obviously the hydrodynamic
interactions have made a large difference.  What is happening can be
partially explained by examining the fluid velocity field
$\mathbf{u}$ and the off-diagonal element of the second 
moment tensor $<\mathbf{pp}>_{xy}$ (related to the
the shear active stresses) in the region between two circular aggregates (see
Fig. \ref{fig:auto_pull3}). Though small in magnitude, the fluid flow
is non-trivial and persistent, and $<\mathbf{pp}>_{xy}$ is close in
magnitude to $<\mathbf{pp}>_{xx}$.  Most importantly, observe that in
the saddle point regions between the circular aggregates the fluid flows are
such that they keep the aggregates apart. By this mechanism, the
hydrodynamic interactions between the aggregates seem to have slowed
down, if not stopped altogether, further aggregation and coarsening.

\subsection{Auto-chemotaxis of Pusher Suspensions}

\begin{figure*}[htps]
\vspace{-0.1in}
\centering
  \includegraphics[width=6in, height=3.5in]{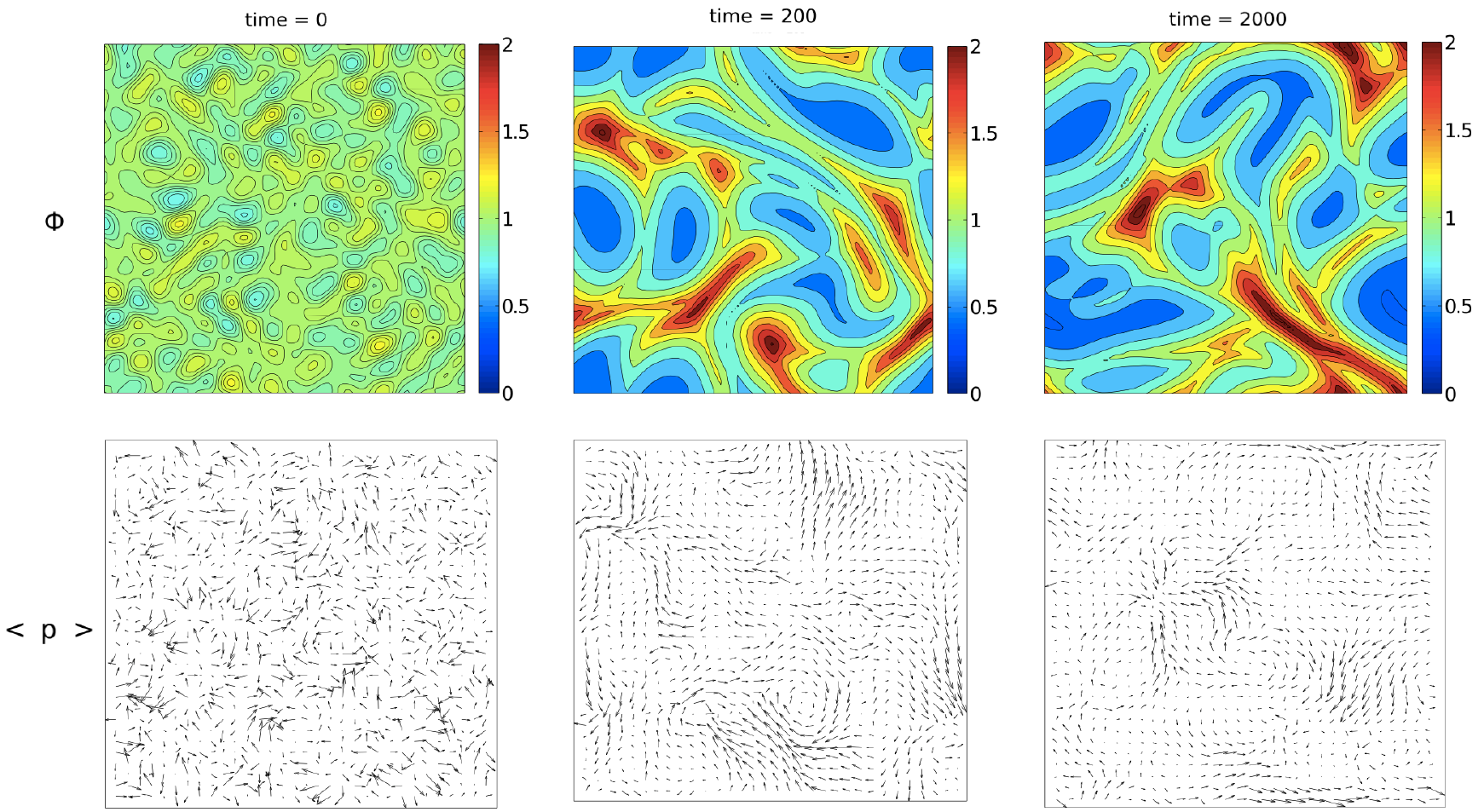}
  \vspace{-0.075in}
\caption{ (Color online) Auto-chemotaxis in a pusher swimmer suspension at times $t=0,200,2000$. The first row shows concentration of swimmers $\Phi$, the second row shows the swimmer mean direction $<\mathbf{p}>$.At $t=$200, $max|<\mathbf{p}>|=0.111$ and at $t=$2000, $max|<\mathbf{p}>|=0.118$} 
\vspace{-0.1in}
 \label{fig:auto_push}
\end{figure*}

\begin{figure*}[htps]
\centering
  \includegraphics[width=5in, height =1.6in]{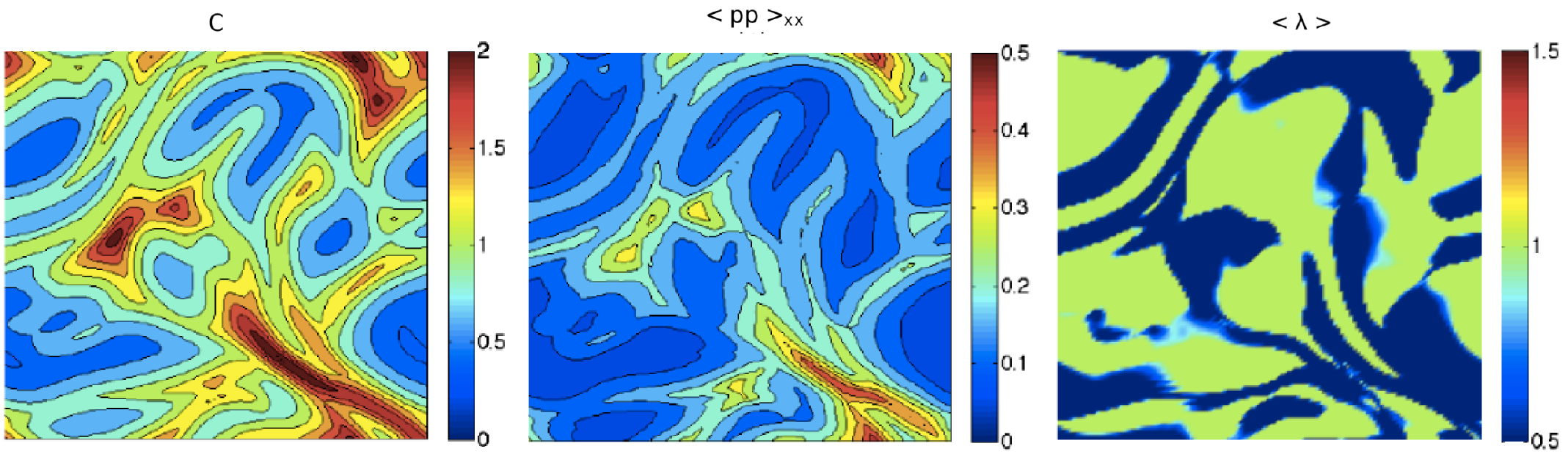}
    \vspace{-0.075in}
\caption{ (Color online) Auto-chemotaxis in a pusher swimmer suspension at time $t=2000$. Shown are the chemoattractant concentration $C$, the diagonal entry $<\mathbf{pp}>_{xx}$
    of the second moment tensor of the distribution function, and the mean tumbling frequency $<\lambda(\mathcal{D}_t C)>$.} 
    \vspace{-0.1in}
 \label{fig:auto_push1}
\end{figure*}

\begin{figure}[htps]
\centering
  \includegraphics[width=3.2in, height=1.6in]{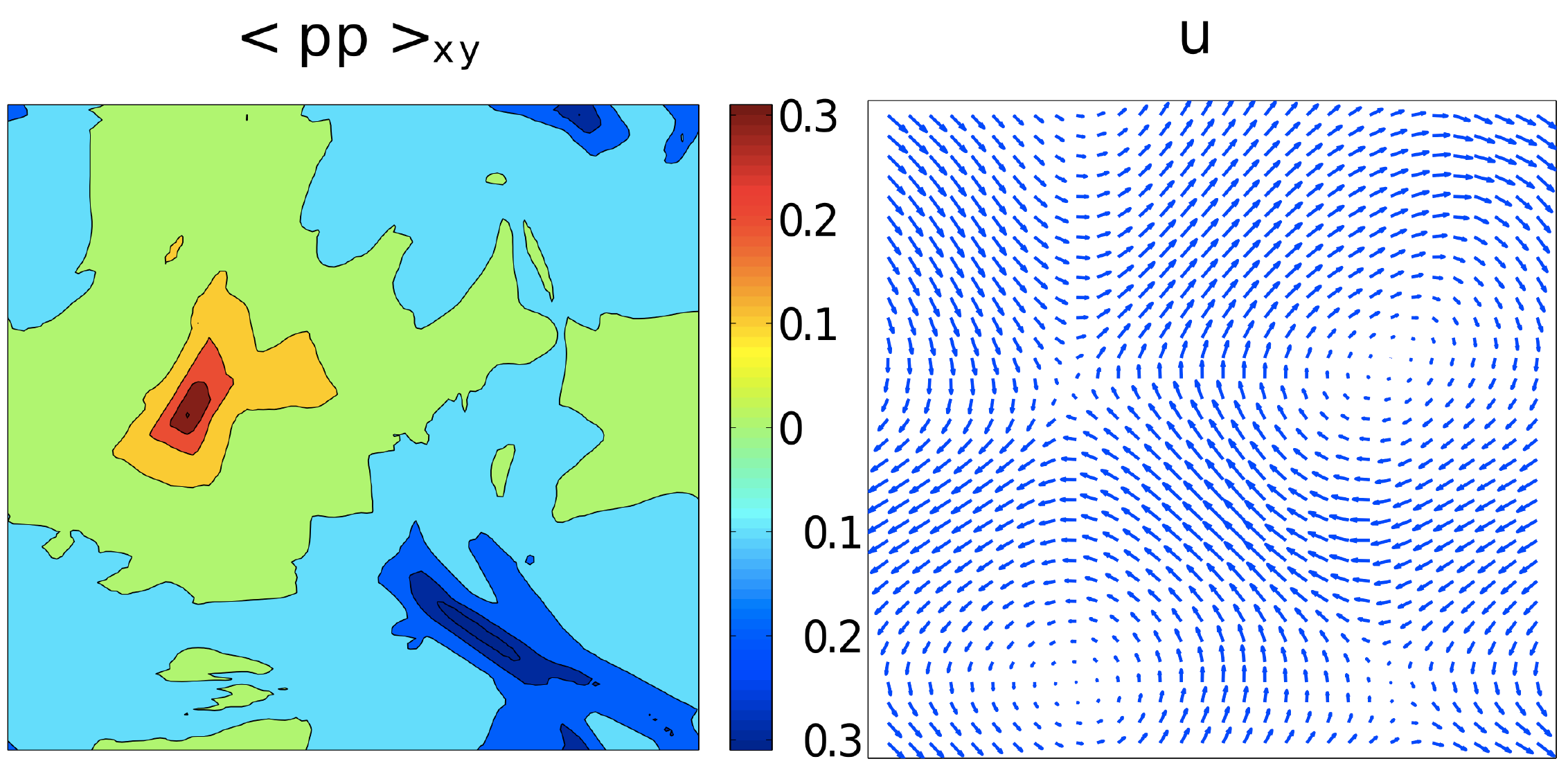}
  \vspace{-0.1in}
  \caption{ (Color online) Auto-chemotactic pusher suspension dynamics at time
    $t=2000$ as shown in Fig. \ref{fig:auto_push}. Left: the off-diagonal entry of the second moment 
    tensor $<\mathbf{pp}>_{xy}$ (related to the active stress). Right: the generated fluid velocity
    $\mathbf{u}$ with $max|\mathbf{u}|=0.87$.}
    \vspace{-0.1in}
 \label{fig:auto_push3}
\end{figure}

We now consider the same numerical experiment for a suspension of
pushers (setting $\alpha=-1$), again keeping all the other parameters
the same -- see the red mark (c) in the phase diagram of 
Fig.~\ref{fig:phasespace}. Linear stability analysis suggests an aggregation
instability due to auto-chemotaxis, and the value of $\lambda_0$ is
large enough to suppress the hydrodynamic linear instability (refer to
Fig.~\ref{fig:GrowthRates}a and shift down the hydrodynamic
instability branches by $\lambda_0=0.25$, making all eigenvalues
negative).

In Figs. \ref{fig:auto_push} and \ref{fig:auto_push1} we see that at
earlier times the dynamics is dominated by aggregation into regions of
high swimmer concentration, as in the neutral swimmer and puller
cases. However, swimmer concentration also leads to locally increased
active stresses which create strong destabilizing fluid flows.
These unsteady fluid flows are significant in magnitude and
macroscopic in scale, and push around the regions of concentrated
swimmers and the chemo-attractant. The resulting dynamics is one of
fragmented regions of aggregation and constant flow instability,
seemingly chaotic in nature. 
The flows have apparently suppressed further growth in the swimmer
concentration.

To reinforce this point, we plot in Fig.~\ref{fig:auto_push3}a the
maximum of the fluid velocity. This shows that the fluid flow is of
significant magnitude, especially considering that this is 
for parameters for which linear
stability analysis predicted no fluid flows at all. The viscous
dissipation $ \mathbf{S} = \int d \mathbf{x} \mathbf{E}: \mathbf{E}$,
which contributes positively to the system configurational entropy in
pusher suspensions (see its significance explained in \cite{SaintShelley08b}),
 is shown in Fig.~\ref{fig:auto_push3}b.

\begin{figure}[htps]
\vspace{-0.1in}
\centering
  \includegraphics[width=2.5in, height=2in]{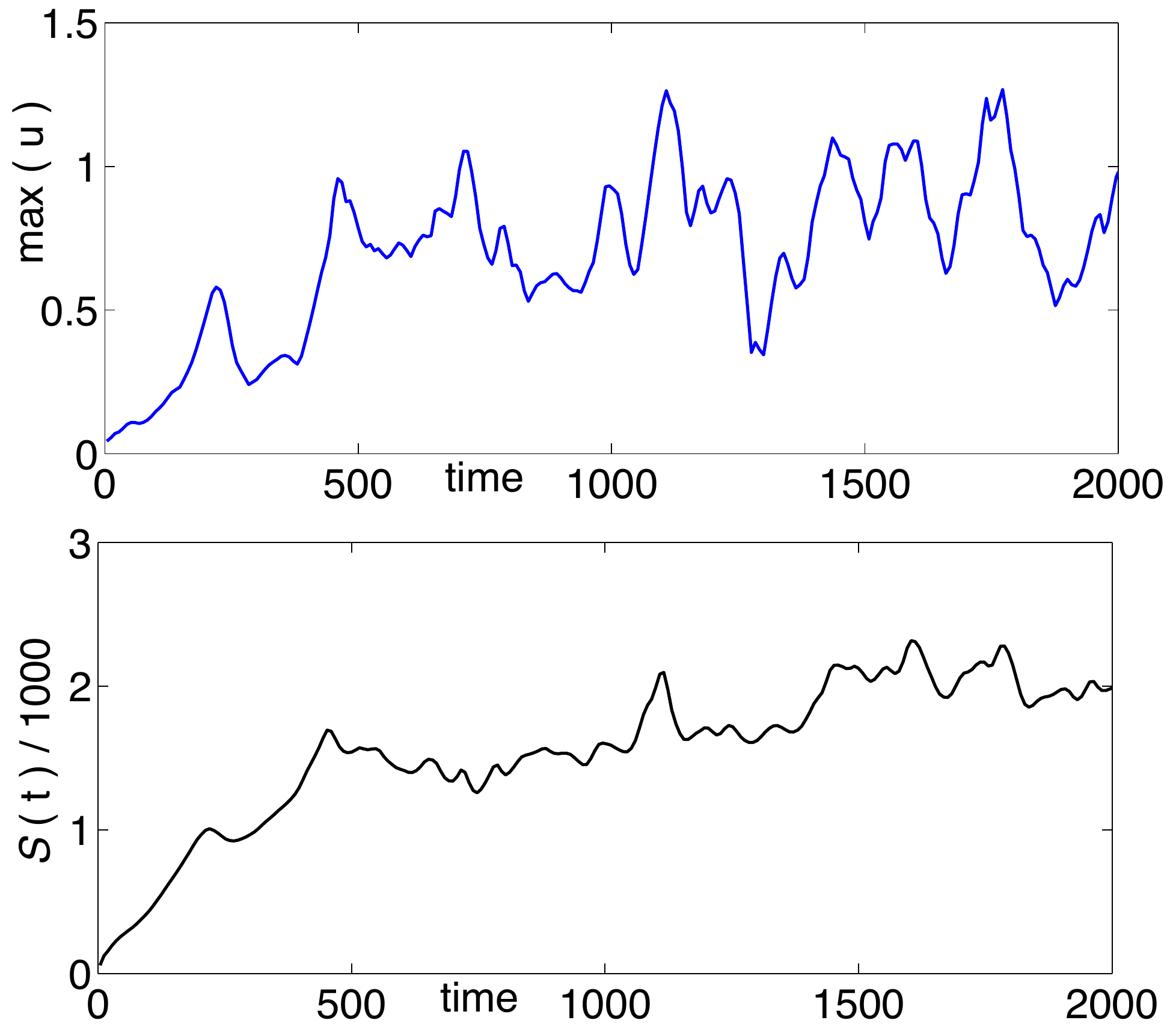}
  \vspace{-0.1in}
 \caption{ (Color online) The maximum fluid speed $max ( |\mathbf{u}| )$ (top) and the configurational entropy $\mathit{S}$ (bottom) vs. time for the auto-chemotactic pusher suspension shown in Fig. \ref{fig:auto_push}.}
 \vspace{-0.1in}
 \label{fig:auto_push3}
\end{figure}

\subsection{Interactions Limit Chemotactic Growth}

\begin{figure}[htps]
\centering
\includegraphics[width=2in, height=2.2in]{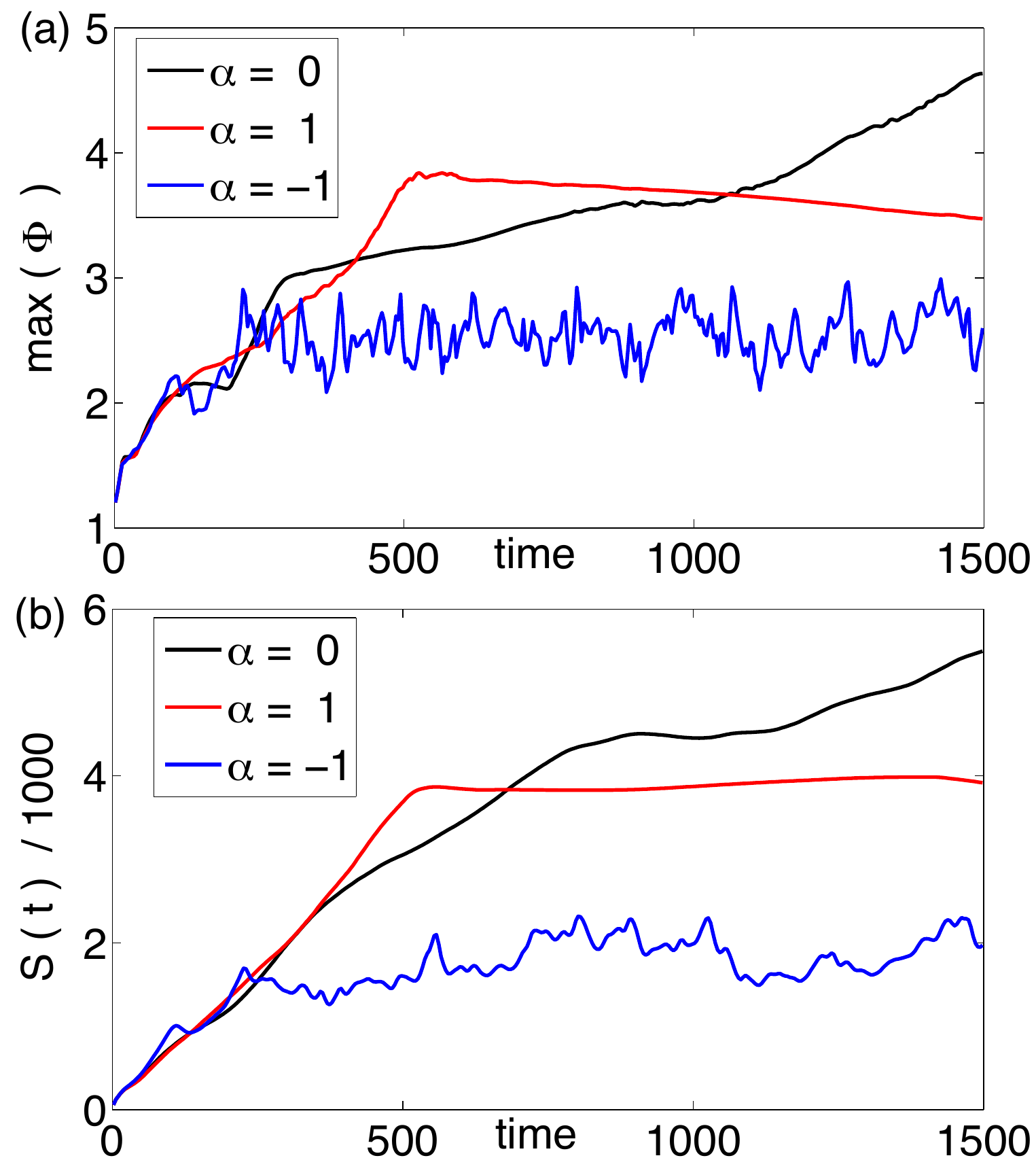}
 \caption{ (Color online) (a) Comparisons in the maximum swimmer concentration $\Phi$, and (b) configurational entropy $\mathit{S}$ in the neutral swimmer ($\alpha=0$, Fig. \ref{fig:auto_neut}), puller ($\alpha=1$, Fig. \ref{fig:auto_pull}) and pusher ($\alpha=-1$, Fig. \ref{fig:auto_push}) suspensions with otherwise the same chemotactic parameters.}
 \label{fig:auto_comparison}
\end{figure}

As noted above, hydrodynamic interactions can alter auto-chemotactic
growth significantly. To quantify this we track the maximum of the
swimmer concentration $\Phi$ and the the configurational entropy
$\mathit{S} = \int d \mathbf{x} \int d \mathbf{p} (\Psi/\Psi_0) \log
(\Psi/\Psi_0) $ in Fig.~\ref{fig:auto_comparison}.  We observe that
these two quantities continually
increase for suspensions of neutral swimmers, but seem to be capped
for the suspensions of pullers or pushers. The chemotactic growth here
is bounded by the hydrodynamical interactions and the fluid flows
the swimmers generate.

\subsection{Mixed Dynamics in Pusher Suspensions}

It is well-known that pusher suspensions ($\alpha<0$) develop a
hydrodynamic instability \cite{SaintShelley08, SaintShelley08b, SubKoch09}. 
In that case without chemotaxis, the nonlinear dynamics
produces concentration bands that stretch and fold in a quasi-periodic
manner, giving rise to strongly mixing fluid flows. 
We now investigate what
happens to the suspension dynamics when auto-chemotaxis is included.  This
corresponds to the ``mixed dynamics'' area in the phase diagram for
pushers in Fig.~\ref{fig:phasespace}a.

For this purpose, we perform nonlinear simulations with $\lambda_0=0.025$,
$\chi=50$, $\beta_1=\beta_2=1/4$, $D_c =0.05$, for which linear theory
predicts dynamics with both strong auto-chemotactic and hydrodynamic
instabilities (see red marker (d) in the phase diagram of Fig. \ref{fig:phasespace}). 
For comparison we include the cases of purely-tumbling
suspensions ($\lambda_0=0.025, \chi=0$), non-chemotactic suspension
($\lambda_0=0$), and another case for which linear analysis predicts
just hydrodynamic, but no auto-chemotactic instability (with
$\lambda_0 \chi \beta_2/\beta_1 <1$). 

\begin{figure}[htps]
\centering
  \includegraphics[width=3.2in]{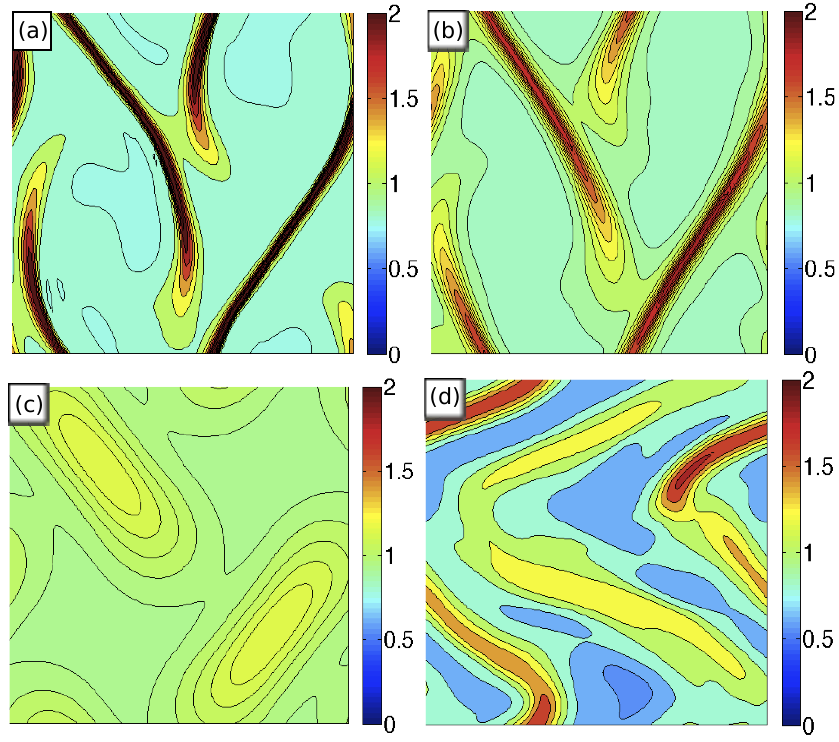}
 \caption{ (color online) Swimmer concentration $\Phi$ in pusher suspensions that are: (a) strongly auto-chemotactic $\lambda_0=0.025$, $\chi=50$, (b) weakly auto-chemotactic $\lambda_0=0.025$, $\chi=35$ ($\lambda_0 \chi \beta_2/\beta_1 <1$), (c) purely tumbling but non-chemotactic $\lambda_0=0.025$, $\chi=0$, and (d) non-tumbling and non-chemotactic $\lambda_0=0$.}
 \label{fig:auto_push4}
\end{figure}

\begin{figure}[htps]
 \centering
\includegraphics[width=2in, height=2.2in]{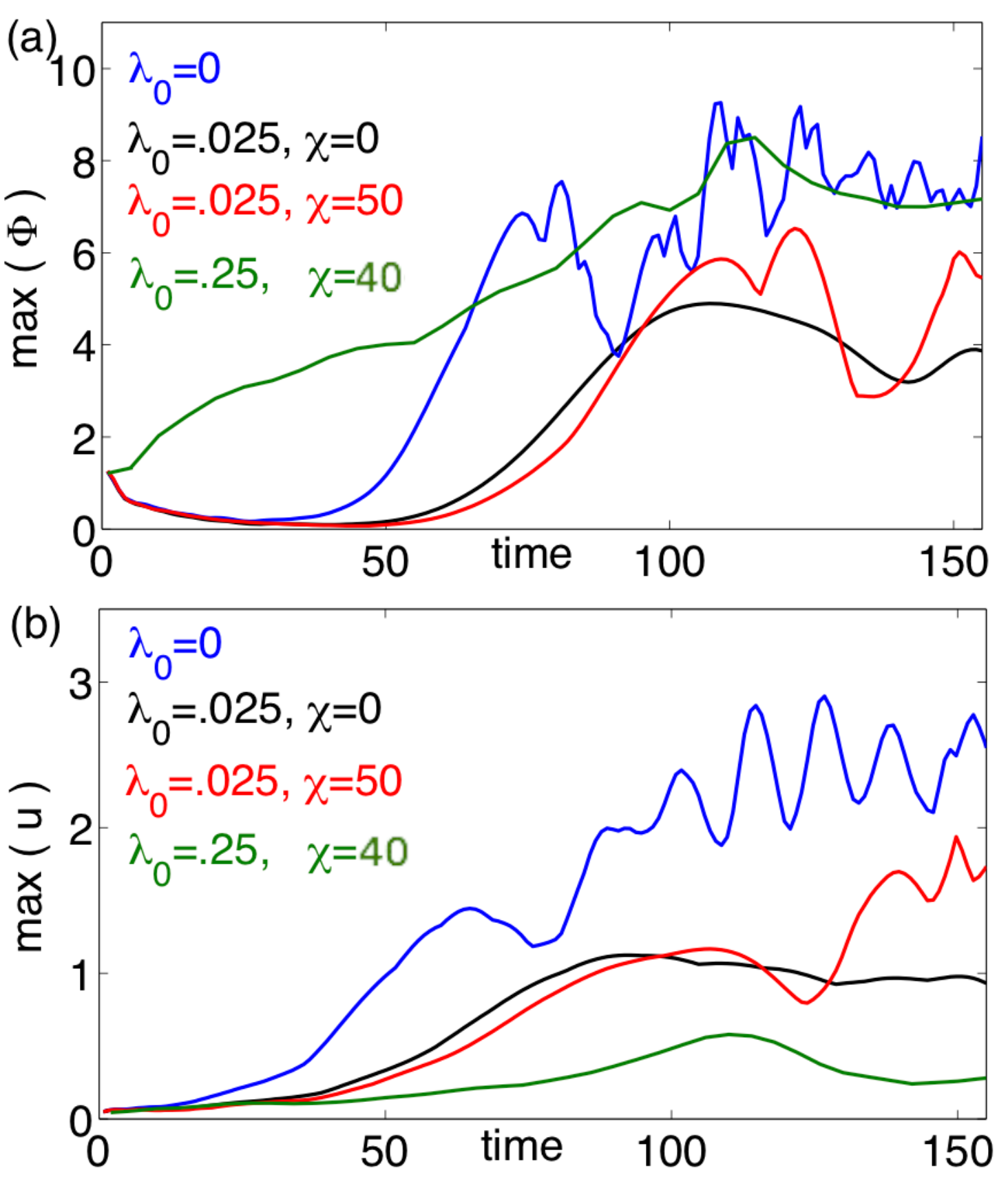}
 \caption{ (color online) Dynamics in pusher suspensions of Figs. ~\ref{fig:auto_push4}a-c: (a) maximum of the swimmer concentration $\Phi$, (b) maximum of the fluid speed $\mathbf{u}$. }
 \label{fig:MixedDyn_Max}
\end{figure}

Fig.~\ref{fig:auto_push4}
shows plots of the swimmers concentration at the onset of the mixing
regime. Auto-chemotactic swimmers produce chemo-attractant as well as
aggregate towards it. A strongly mixing flow emerges and advects both
swimmers and chemo-attractant. The chemo-attractant dynamics closely
follows those of the swimmer concentration, resulting in 
\textit{dynamic aggregation} of swimmers occurring due to the local
auto-chemotactic tendency. This effect is seen from the sharper and
narrower concentration bands in the auto-chemotactic suspension in
Fig.~\ref{fig:auto_push4}a compared to non-chemotactic tumblers in
Fig.~\ref{fig:auto_push4}c. Hence auto-chemotaxis stabilizes the
formation of concentration bands that pure tumbling had diminished through its
diffusion-like effect. The effect is apparent even for the case where
no auto-chemotactic instability is predicted by linear theory
($\lambda_0 \chi \beta_2/\beta_1 <1$), as shown in
Fig.~\ref{fig:auto_push4}b. In Fig.~\ref{fig:auto_push4}a-d we see
that auto-chemotaxis has also hastened the onset of the mixing regime
when compared to the purely-tumbling pusher suspension. Linear
stability predicts that pure tumbling has a stabilizing effect on the
suspension. This is confirmed in simulations when comparing
the weak concentration bands for pure-tumbler compared to the
non-tumbling suspension of non-chemotactic pushers in
Fig.~\ref{fig:auto_push4}cd. These effects are also illustrated in
plots of the swimmer concentration and generated fluid flow in
Fig. \ref{fig:MixedDyn_Max}.

\subsection{Similarities between the Chemotaxis Models}
\begin{figure}[htps]
\centering
  \includegraphics[width=3.2in]{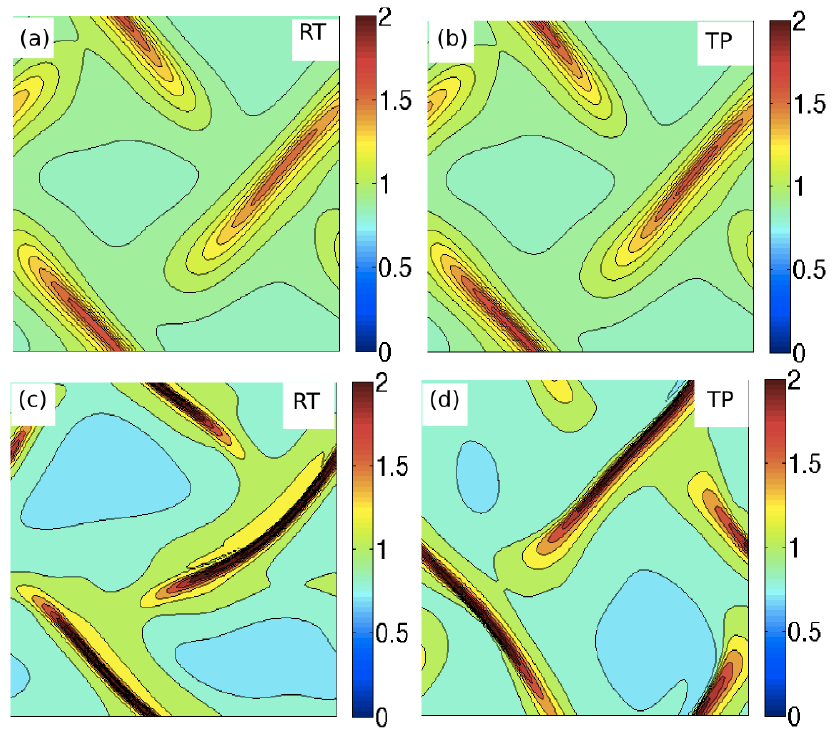}
   \vspace{-.15in}
 \caption{ (color online) Swimmer concentration $\Phi$ in pusher suspensions for the RT chemotaxis model with $\lambda_0=0.025, \chi=60, d_r=0.025$ (left) and TP chemotaxis model with $\xi = 0.75, d_r=0.175$ (right) at times $t=100$ (top) and $t=150$ (bottom). }
 \label{fig:comp_models}
 \vspace{-.15in}
\end{figure}

We illustrate the qualitative similarities in the dynamics of the two
chemotaxis models when the parameters are matched as suggested by the
linear theory: $\lambda_0 \approx 6 d_r$ and $ \xi \approx \lambda_0
\chi/2 $. Fig.~\ref{fig:comp_models} shows pusher swimmer
concentration for the two models at the onset of mixing. The profiles
and dynamics are remarkably similar, and such similarity is also
observed in plots of chemo-attractant, the generated fluid flows, and
the normal and shear stresses (not shown).

The Turning-Particle model assumes that a swimmer is able to detect
the local chemo-attractant gradient and adjust its orientation to swim
towards the regions of high chemo-attractant concentration. This
chemotactic response is induced through a torque that aligns the
swimmers with the chemo-attractant gradient. While the model is not
applicable to bacteria, it is interesting that there are connections
to the Run-and-Tumble chemotaxis model in the linear analysis and the
also the nonlinear dynamics in the long wave regimes. This connection
might be helpful in designing models for direct swimmer simulations of
chemotactic response. On that note, approaches that model chemotaxis
as a bias in the swimmer direction have been applied in other studies
of individual swimmer models
\cite{DillonEtAl95,HopkinsFauci02,TaktikosEtAl12}.

\section{Discussion and Conclusion}
\label{Conclusion}

We have presented and elaborated upon two kinetic models that couple
chemo-attractant production and response in colonies of micro-swimmers
with the fluid flows that the swimmers collectively generate by their
motion.  These models, and our study of them, merge together two
separate areas of investigation: chemotactic aggregation due to
population-produced chemo-attractants, and the hydrodynamics of active
motile suspensions.  In classical models of attractive chemotaxis,
concentration through aggregation is often unbounded and can blow up
in finite time, and such behaviors are sometimes avoided in extended
models through the inclusion of {\it ad hoc} saturation terms.  We
show here that the flows generated by motile suspensions can also
limit aggregation growth, though for very different reasons depending
upon the swimmer type. That the collective flows associated with
motility can achieve this has not been previously demonstrated.

Still, the application of our modeling may be limited. Our active
suspension model is a dilute to semi-dilute theory that does not
include local interactions between swimmers, either hydrodynamic or
steric; see
\cite{CisnEtAl11,DrescherEtAl11,Dunkel13,Hensink12,SokolovAranson12}
for relevant experiments using bacteria. In denser suspensions the
swimmer size limits local swimmer density through steric interactions,
and well-founded models that combine these with hydrodynamic
interactions are in development (for one recent attempt, see
\cite{Ezhilan13}). We do note a recent study by Taktikos {\em et. al.}
\cite{TaktikosEtAl12} for discrete disk-shaped chemotactic random
walkers in 2D with steric but no hydrodynamic effects showing that steric
interactions can limit aggregation, as indeed they must. Further, in
dense suspensions it is not clear how run-and-tumble dynamics, as has
been observed and modeled, is affected by crowding and steric
interactions. Intuitively, one expects the swimmer tumbling frequency
to decrease in denser suspensions where mobility is limited due to
crowding.

While these theoretical results on the coupling of auto-chemotaxis
with collectively-generated flows have not yet been systematically
studied in an experimental setting, this might be possible with the
specific engineering of the dynamics of locomotion and
chemosensing \cite{ParkEtAl03}. Moreover, the interplay between
locomotion, fluid flows, chemotaxis and quorum sensing can be further
illuminated through the controlled introduction of exogenous
chemo-attractants \cite{LiuEtAl11}. As a relevant example, a recent
experiment by Saragosti et. al. \cite{SaragostiEtAl11} used exogenous
chemo-attractants in conjunction with those produced by swimming {\it
  E. coli} to induce aggregation and traveling waves of bacteria in a
channel. As side-note on this, in our first study (Lushi {\it et.
  al.} \cite{LGS12}) we investigated the RT model using parameters
close to those of the Saragosti {\it et al} \cite{SaragostiEtAl11} experiment and found the
production of filamentary aggregates (see Fig.~1(iv) and Supplementary
Material of \cite{LGS12}). Despite this system being well outside of
the regime of hydrodynamic instability (as predicted by our linear
analysis), hydrodynamics was plainly important in their local
dynamics. Perhaps in a setting that also included the effects of
confinement and steric ordering, these aggregates would transition to
traveling waves. Obviously this model is easily modified to study the
effects of external chemo-attractants (see, for example,
\cite{Ezhilan12, LushiDissert}).  Chemotaxis in bacterial colonies has
been previously exploited for enhancing mixing in microfluidic devices
\cite{KimBreuer07}, but it has not yet been studied experimentally how
the mixing process might be affected by auto-chemotaxis. Lastly,
chemotactic-like behavior is also observed in suspensions of synthetic
micro-swimmers that exist in micro-fluidic environments (see
\cite{HongEtAl07} for experiments, and \cite{TBZS2013,MWAP2013} for
recent theory). Such chemotactic responses might be exploited in the
future in technological applications
\cite{HongEtAl07,Kagan10,SenVel09}.

\begin{acknowledgments}
The work was supported in part by the NSF grants DMS-0652775, DMS-0652795, DOE grant DEFG02-00ER25053 (E.L., M.J.S), a Dean's Dissertation Fellowship (E.L.), and an ERC Advanced Investigator Grant No. 247333 (R.E.G.). 
\end{acknowledgments}

%

\end{document}